\documentclass[ejs, preprint]{imsart}
\usepackage{amsthm}
\usepackage{mathtools}
\usepackage{tikz}
\usepackage{amsmath}
\usepackage{amsfonts}
\usepackage{bbm}
\usepackage{natbib}
\usepackage[colorlinks,citecolor=blue,urlcolor=blue,filecolor=blue,backref=page]{hyperref}
\usepackage{graphicx}
\usepackage{float}
\usepackage{booktabs}
\usepackage{threeparttable}
\usepackage{multirow}

\startlocaldefs
\graphicspath{ {./images/} }
\newcommand{\rsq}{R^2}
\newcommand{\cov}{\text{cov}}

\newcommand{\var}{\text{var}}
\newcommand{\mbe}{\mathbb{E}}

\newcommand{\mcs}{\mathcal{S}}
\newcommand{\given}{\, | \,}

\newcommand{\alr}{\text{alr}}

\newcommand{\KL}[2]{\text{KL}\left(  #1 ||  #2 \right)}

\newcommand{\normal}{\mathcal{N}}
\newcommand{\logisticnormal}{\text{LogisticNormal}}
\newcommand{\LN}{\text{LogisticNormal}}
\newcommand{\dirichlet}{\text{Dirichlet}}

\newcommand{\betadist}{\text{Beta}}
\newcommand{\betaprime}{\text{BetaPrime}}
\newcommand{\gammadist}{\text{Gamma}}

\newcommand{\gigdist}{\text{GenInvGaussian}}
\newcommand{\invgammadist}{\text{InvGamma}}

\DeclareMathOperator*{\argmin}{arg\,min} 
\newcommand{\rmse}{\text{RMSE}}

\newcommand{\elpd}{\text{ELPD}}

\newcommand{\myosfresults}{\url{https://osf.io/ns2cv/}}

\usetikzlibrary{shapes,fit} %use shapes library if you need ellipse
\usetikzlibrary{bayesnet}
\usetikzlibrary{shapes,decorations,arrows,calc,arrows.meta,fit,positioning}
\tikzset{
    -Latex,auto,node distance =1 cm and 1 cm,semithick,
    state/.style ={circle, draw, minimum width = 1.0 cm},
    point/.style = {circle, draw, inner sep=0.04cm,fill,node contents={}}, 
    vmissing/.style={
    draw=none, 
    scale=1,
    text height=0.111cm,
    execute at begin node=\color{black}$\vdots$
    },
    hmissing/.style={
    draw=none, 
    scale=1,
    text height=0.111cm,
    execute at begin node=\color{black}$...$
    },
    bidirected/.style={Latex-Latex,dashed},
    el/.style = {inner sep=2pt, align=left, sloped}
}

\endlocaldefs

\newcommand*{\getFirstNameFirstAuthor}{Javier Enrique}
\newcommand*{\getLastNameFirstAuthor}{Aguilar}

\newcommand*{\getFirstNameSecondAuthor}{Paul-Christian}
\newcommand*{\getLastNameSecondAuthor}{Bürkner}

\newcommand*{\getRunAuthor}{J.E. Aguilar and P. Bürkner }

\newcommand*{\getTitle}{Generalized Decomposition Priors on R2}
\newcommand*{\getRunningTitle}{Generalized Decomposition Priors on R2}

\newcommand*{\getFirstAddress}{Department of Statistics, TU Dortmund University, Germany}

\newcommand*{\getSecondAddress}{Cluster of Excellence SimTech, University of Stuttgart, Germany}

\newcommand*{\getMailFirstAuthor}{javier.aguilarr@icloud.com}
\newcommand*{\getMailSecondAuthor}{paul.buerkner@gmail.com}

\newcommand*{\getURLFirstAuthor}{https://jear2412.github.io}
\newcommand*{\getURLSecondAuthor}{https://paul-buerkner.github.io}

\pubyear{2025}
\volume{TBA}
\issue{TBA}
\firstpage{1}
\lastpage{1}

\begin{document}

%% *** Frontmatter *** 
\begin{frontmatter}

\title{ \getTitle  }
\runtitle{ \getRunningTitle}
\date{07.01.25}

\begin{aug}
\author{\fnms{\getFirstNameFirstAuthor} \snm{\getLastNameFirstAuthor}\thanksref{addr1,addr2,t1}\ead[label=e1]{\getMailFirstAuthor}%
\ead[label=u1,url]{\getURLFirstAuthor}
}
\and
\author{\fnms{\getFirstNameSecondAuthor} \snm{\getLastNameSecondAuthor}\thanksref{addr1,t2}\ead[label=e2]{\getMailSecondAuthor}%
\ead[label=u2,url]{\getURLSecondAuthor}}

\runauthor{ \getRunAuthor}

\address[addr1]{\getFirstAddress}
\address[addr2]{\getSecondAddress}

\thankstext{t1}{Corresponding author: Javier Enrique Aguilar \printead{e1} \printead{u1}}

\thankstext{t2}{\printead{e2} \printead{u2}}

\end{aug}

\begin{abstract}
\noindent
%--- Old
% The adoption of continuous shrinkage priors in high-dimensional linear models has gained momentum, driven by their theoretical and practical advantages. One of these shrinkage priors is the R2D2 prior, which comes with intuitive hyperparameters and well understood theoretical properties.  
% The core idea is to specify a prior on the percentage of explained variance $R^2$ and to conduct a Dirichlet decomposition to distribute the explained variance among all the regression terms of the model.  Due to the properties of the Dirichlet distribution, the competition among variance components tends to gravitate towards negative dependence structures, fully determined by the individual components' means. Yet, in reality, specific coefficients or groups may compete differently for the total variability than the Dirichlet would allow for. In this work we address this limitation by proposing a generalization of the R2D2 prior, which we term the \textit{Generalized Decomposition R2} (GDR2) prior. 

% Our new prior provides great flexibility in expressing dependency structures as well as enhanced shrinkage properties. Specifically, we explore the capabilities of variance decomposition via logistic normal distributions.
% Through extensive simulations and real-world case studies, we demonstrate that GDR2 priors yield strongly improved out-of-sample predictive performance and parameter recovery compared to R2D2 priors with similar hyper-parameter choices.
%--- New 
The adoption of continuous shrinkage priors in high-dimensional linear models has gained widespread attention due to their practical and theoretical advantages. Among them, the R2D2 prior has gained popularity for its intuitive specification of the proportion of explained variance ($R^2$) and its theoretically grounded properties. The R2D2 prior allocates variance among regression terms through a Dirichlet decomposition.  However, this approach inherently limits the dependency structure among variance components to the negative dependence modeled by the Dirichlet distribution, which is fully determined by the mean.  This limitation hinders the prior’s ability to capture more nuanced or positive dependency patterns that may arise in real-world data.

To address this, we propose the Generalized Decomposition R2 (GDR2) prior, which replaces the Dirichlet decomposition with the more flexible Logistic-Normal distribution and its variants. By allowing richer dependency structures, the GDR2 prior accommodates more realistic and adaptable competition among variance components, enhancing the expressiveness and applicability of $R^2$-based priors in practice. Through simulations and real-world benchmarks, we demonstrate that the GDR2 prior improves out-of-sample predictive performance and parameter recovery compared to the R2D2 prior. Our framework bridges the gap between flexibility in variance decomposition and practical implementation, advancing the utility of shrinkage priors in complex regression settings.
\end{abstract}

%% ** Keywords **

%\tableofcontents
\end{frontmatter}

% ** Main content **

\section{Introduction}

Linear regression is one of the most widely used statistical techniques, serving as the foundation for many advanced modeling methods \citep{gelman_hill_2006, gelman2013bda}. However, its limitations become apparent in high-dimensional settings or when predictors exhibit multicollinearity \citep{Ridge, tibshirani1996lasso, giraud2014HD}. In such cases, parameter estimation becomes unstable, and traditional models often fail to deliver reliable or interpretable results. To address these challenges, a variety of priors with exceptional theoretical and empirical properties have been developed over the last decade. These priors often assume that the true underlying model is sparse, meaning that many regression coefficients are exactly zero, and have the purpose of enforcing sparsity \citep{Horseshoe, vanDP2016Conditions, vanDP2021theoretical}.

Even in cases where the true model is not sparse, sparse approximations have been widely recommended and successfully applied across a range of disciplines \citep{tibshirani1996lasso, giraud2014HD, vanderpasblackhs,  hastie2015statisticalsparsity, ZhangCredibleRegions, PolsonHSmeetsLasso}. Dense solutions, while theoretically possible, tend to violate the principle of parsimony by increasing model complexity, which can lead to overfitting, reduced interpretability, and higher computational costs \citep{giraud2014HD, hastie2015statisticalsparsity, BayesianHDComplexity}. As a result, there has been significant interest in priors that balance regularization and flexibility while remaining computationally efficient.

Continuous global-local shrinkage priors are a powerful class of priors that strike a balance between sparsity and regularization while maintaining flexibility through the incorporation of both global and local shrinkage effects \citep{bayesian_variable_selection_handbook}. Over the past decade, these priors have been extensively studied and widely adopted within the Bayesian community due to their excellent empirical performance, computational efficiency, and the availability of fast implementations \citep{vanDP2016Conditions, PiironenHorseshoe, vanDP2017UQHorseshoe, ScalableHSBhattacharya}.

Furthermore, a broad class of continuous global-local shrinkage priors has been shown to achieve near-minimax recovery rates \citep{ghosalconvergence, vanderpasblackhs, vanDP2016Conditions, bayesian_variable_selection_handbook}. General conditions over the local scales have also been established to ensure that the posterior distribution concentrates at the minimax estimation rate under these shrinkage priors \citep{vanDP2016Conditions, Rockov2018BayesianEO, PolsonHSmeetsLasso}. These theoretical guarantees provide a foundation for understanding why the priors commonly used in practice perform so well: the observed empirical success is, in essence, a reflection of the strong theoretical properties underpinning these models.

Prominent examples of shrinkage priors include the horseshoe, horseshoe-plus, three-parameter beta, normal gamma, and generalized double Pareto priors \citep{Horseshoe, Horseshoeplus, griffin2010inference, armagan2011generalized, armagan2013generalized}. These priors typically assume the local scales are conditionally independent given the global scale. However, an alternative family of priors introduces dependencies between the scales via joint distributions, albeit in a much smaller subset of the literature. Examples include the Dirichlet-Laplace, R2D2, and its multilevel extension, the R2D2M2 prior \citep{DirichletLaplace, r2d2zhang, aguilar_intuitive_2023}. These approaches model local scales as proportions using a Dirichlet distribution. Therefore the competition among variance components will only be able to gravitate towards negative dependence structures fully specified by the mean. Yet, in reality, specific coefficients or groups may compete differently for the total variability than the Dirichlet would allow for. Thus, the ability to capture more nuanced or positive dependence structures among variance components, which might better reflect real-world scenarios is severely limited. 

The value of moving beyond Dirichlet distributions to more flexible alternatives has long been recognized in fields such as Compositional Data Analysis \cite{AitchisonMonograph, Aitchison40years}, Categorical Data Analysis \citep{CATDAagresti_bayesian_2005}, and Machine Learning in areas such as Correlated Topic Modeling \citep{CorTM, ScalableCTMLN}. 
Despite their success, these methods have not been explored in the context of shrinkage priors, leaving a gap in the literature.

Among existing shrinkage priors, the R2D2 prior is particularly notable for its focus on global quantities of interest—such as the proportion of explained variance ($R^2$)—and its ability to jointly regularize regression coefficients \citep{r2d2zhang, aguilar_intuitive_2023, prioreli}. By simplifying prior elicitation and enhancing interpretability, the R2D2 framework has become a valuable tool for practitioners. However, its reliance on Dirichlet decompositions limits its flexibility in modeling dependencies among variance components.

The R2D2 prior specifies a prior on $R^2$ and uses a Dirichlet decomposition to allocate the variance across regression terms. In this work, we address the limitations of the Dirichlet-based decomposition by introducing the Generalized Decomposition R2 (GDR2) prior, which extends the R2D2 framework using Logistic-Normal decompositions and their variants. This extension enables richer, more expressive parameterizations of dependency structures among variance components, capturing complex relationships that are inaccessible with Dirichlet-based approaches. By combining the intuitive interpretability of the R2D2 framework with increased flexibility, the GDR2 prior represents a significant step forward in the development of continuous global-local shrinkage priors.

Our work is organized as follows: we begin by discussing preliminaries and the implied distributions on the proportions of variance and $R^2$ by shrinkage priors in Sections \ref{sec:prelims}, \ref{subsec:implicitpriors} respectively. We proceed in Section \ref{sec:genr2} by introduing and describing the GDR2 prior framework for high-dimensional Bayesian linear regression. To circumvent the limitations associated with the Dirichlet decomposition, we propose the use of Logistic-Normal decompositions and its variants as an alternative in the decomposition step of $R^2$-based shrinkage priors in Section \ref{sec:simplexdists}. Section \ref{sec:hyperparam_spec} follows with a discussion on hyperparameter specification, offering intuitive explanations for selecting hyperparameters and exploring their effects on shrinkage. We believe this contribution is particularly valuable, as hyperparameter specification for Logistic-Normal distributions is rarely addressed in the literature.

In Section \ref{sec:experiments}, we present the results of simulation studies and real life experiments conducted to evaluate the capabilities of the GDR2 prior. Our findings demonstrate that allowing for richer dependency structures among variance components improves out-of-sample predictive performance and parameter recovery. For these simulations and experiments, we implemented our models in the probabilistic programming language Stan \citep{StanJSS, stan2022}, leveraging a parameterization optimized for fast convergence and efficient Hamiltonian Monte Carlo sampling. We also provide a Slice-within-Gibbs sampler for alternative usage in the Appendix \ref{appendix}. To validate our approach, we tested the GDR2 prior on three real-world benchmarks commonly used in the shrinkage prior literature. Across all scenarios, the GDR2 prior consistently led to significantly improved results. All code and replication materials are openly available on the Open Science Framework (\myosfresults).

\section{Methods}

\subsection{Preliminaries}
\label{sec:prelims}

Consider the linear regression model 
%----------------
\begin{align}
    \label{eqn::linreg}
    y_n = x_n' b+ \varepsilon_n, \;  n=1,...,N,
\end{align}
%----------------
where $y_n$ is the $n$th response value, $x_n$ is the $K$ dimensional vector of covariates for the $n$th observation, $b = (b_1,..., b_K)'$ is the vector of regression coefficients, and  $\varepsilon_n \sim \normal(0, \sigma^2)$ is the residual error, with $\sigma$ being the residual standard deviation. 

Continuous Global-Local (GL) shrinkage priors  \citep{ vanDP2016Conditions, BayesPenalizedRegSara, vanDP2021theoretical} are a special type of continuous prior distributions that arise from scale mixtures of normals \citep{mwestScaleNormals}. They take on the form 
%------------------
    \begin{align}
    \label{eqn::glprior}
        b_k \given \lambda_k^2, \tau^2, \sigma^2  \sim \normal \left( 0, \sigma^2 \lambda_k^2 \tau^2\right), \ \ \lambda_k \sim \pi(\lambda_k), k=1,..., K , \ \ \tau \sim \pi(\tau), \sigma \sim \pi(\sigma),
    \end{align}
%------------------
    where $\lambda_k$ represents local scales unique to each regression coefficient $b_k$, $\tau$ denotes a global scale shared across all coefficients, and $\pi(\cdot)$ represents the corresponding (hyper) prior distributions.  Priors on $b$ of the form \eqref{eqn::glprior} centered at zero are designed to shrink each coefficient towards zero thus favouring more sparse solutions. These priors have demonstrated empirical success among practitioners, and their theoretical foundations justify their widespread application  \citep{Ghosh2019, ScalableHSBhattacharya, FOLLETT2019130, Kohns2020HorseshoePB, bayesian_variable_selection_handbook}.
    
    The choice of $\pi(\lambda_k)$ and $\pi(\tau)$ produces different GL shrinkage priors and will have an important effect on their properties. Popular shrinkage priors include the Horseshoe \citep{Horseshoe}, the Normal-Gamma \citep{griffin2010inference},  Generalized Double Pareto \citep{armagan2013generalized}, the Dirichlet-Laplace \citep{DirichletLaplace}, the Regularized Horseshoe \citep{PiironenHorseshoe},  and the R2D2 prior \citep{r2d2zhang}. The global scale $\tau$ controls the overall sparsity, ideally represent the proportion of true signals \citep{PiironenHorseshoe, BayesPenalizedRegSara, vanDP2021theoretical}. The local scales $\lambda_k$ can either counteract or enforce the shrinkage towards zero.
    
    %--------- Shrinkage factors
    The amount of shrinkage exerted on each coefficient towards zero can be quantified by studying the posterior distribution of the coefficients $b$ given the scales  $\lambda_k, \tau, \sigma$ and the observations $y$. The conditional posterior of the regression coefficients $b$ is $b \given y, \lambda, \tau, \sigma \sim \normal \left( \bar{b} , \Sigma_b \right)$, with mean $\bar{b}= \mbe \left( b \given y,  \lambda, \tau, \sigma \right)=  \left(  X'X + 1/\tau^2 \Lambda^{-1} \right)^{-1} X' y $ and and covariance matrix $\Sigma_b= \left(  X'X + 1/\tau^2\Lambda^{-1} \right)^{-1}$, where $\Lambda= \{ \lambda_1^2,..., \lambda_K^2 \}$ is the diagonal matrix containing the local scales. If $X$ is of full rank, then the conditional posterior mean can be expressed as $\bar{b}= \tau^2 \Lambda  \left(\tau^2 \Lambda+ (X'X)^{-1} \right) \hat{b}$ where $\hat{b}$ is the Maximum Likelihood Estimator (MLE). This representation highlights that the conditional posterior mean is a shrunken version of the usual MLE estimate.
    
    For illustration, consider $X=I$, i.e. the normal means problem \citep{stein_estimation_1981,vanDP2016Conditions, Horseshoeplus}. In this case $\bar{b}_k =  \left( 1- \frac{1}{1+\lambda_k^2 \tau^2} \right)\hat{b}_k= (1- \kappa_k) y_k$, where $\kappa_k \coloneqq \frac{1}{1+\lambda_k^2 \tau^2}$. The quantity $\kappa_k$ serves as a \textit{shrinkage factor} and allows us to assess how much shrinkage the prior exerts on the maximum likelihood estimator of $b_k$ \citep{polson2013bayesian, BaiHypothesisNB, PolsonHSmeetsLasso,  aguilar_intuitive_2023}. Importantly, by an application of Fubini's theorem, we can show that $\mbe[b_k \given y_k ] = \left( 1- \mbe[\kappa_k| y_k] \right) y_k$,  demonstrating that the posterior mean of $b_k$ is at most $y_k$ \citep{Horseshoe, TweedieEfron}. 
   
    The prior on $\kappa_k$ is determined by the interplay between the priors on $\lambda_k$ and $\tau$. Thus, calibrating shrinkage requires careful consideration of these priors. Notably, for a fixed value of $\tau$, the definition of $\kappa_k$ reveals that the posterior mean of $b_k$ is influenced by the value of the local scale $\lambda_k$. Small values of $\lambda_k$ will shrink $b_k$ towards zero, whereas large values push the posterior mean of $b_k$ towards the observation $y_k$. If either $\tau \to \infty$ or $\lambda_k \to \infty$ then the MLE is recovered.
    
    In general, a good shrinkage prior should possess the following characteristics \citep{Castillo, PiironenHorseshoe, BayesPenalizedRegSara, PolsonHSmeetsLasso, vanDP2021theoretical}: 1) \textit{Heavy tails:} Sufficient mass in the tails of the prior is crucial to properly recover signals (i.e., truly nonzero coefficients). 2) \textit{Sufficient mass near zero:} Shrinkage priors should allocate enough prior mass near zero in order to shrink redundant (truly zero) coefficients towards it. 3) \textit{Efficient and stable sampling:} While theoretical properties are important, designing shrinkage priors should also consider efficient sampling from the posterior distribution. 
    
    Continuous GL shrinkage priors have gained popularity due to their capability to discern noise from signals, while yielding solutions that avoid a search over the entire space of models. They effectively shrink the influence of covariates considered unimportant towards zero while recovering the values of signals. This holds true even in scenarios where the true vector of regression coefficients $b$ is ultra sparse \citep{Horseshoeplus,Song2017NearlyOB, vanDP2017UQHorseshoe}. Since continuous GL shrinkage priors cannot product \textit{exact} zero estimates, additional posterior variable selection is required to induce sparsity in the posterior estimates, leading to a two step procedure: first perform inference of the GL shrunken model and second employ a decision rule to perform variable selection \citep{BaiHypothesisNB, PiironenProjInf,  vanDP2017UQHorseshoe, ZhangCredibleRegions, pavone2020using}.

\subsection{Implied priors by the local and global scales}
\label{subsec:implicitpriors}

Consider the linear regression model \eqref{eqn::linreg}, where the covariates are standardized, such that $\mathbb{E}(x) = 0$ and $\var(x) = \Sigma_X$, with $\Sigma_X$ having a diagonal of ones. Assume a prior distribution for the regression coefficients $b$ such that $\mathbb{E}[b] = 0$ and $\var(b) = \sigma^2 \Lambda$, where $\Lambda$ is a diagonal matrix with entries $\lambda_1^2, \dots, \lambda_K^2$. The conditional variance of the linear predictor $x'b$ is given by \citep{gelmanregstories2020}:
%----------------------------
\begin{align}
\label{eqn::varlinpred}
    \var(\mu)&=  \sigma^2 \sum_{k=1}^K \lambda_k^2.
\end{align}
%----------------------------
We define the quantity $\omega^2 \coloneqq \sum_{k=1}^K \lambda_k^2$ as the \textit{total variance}. Equation \eqref{eqn::varlinpred} provides insight into the prior distribution for the proportion of explained variance, $R^2$ \citep{gelmanregstories2020}, induced by a GL shrinkage prior. The proportion of explained variance is defined as the square of the correlation coefficient between $y$ and the linear predictor $x’b$:
%----------------------------
\begin{align}
    \label{eq::r2def}
    R^2 \coloneqq \text{corr}^2(y, x'b)= \frac{\var(x'b)}{\var(x'b)+\sigma^2}=\frac{\omega^2 }{\omega^2+1}.
\end{align}
%----------------------------
Thus, there is a one-to-one relationship between the priors set on $R^2$ and those implied on $\omega^2$. Specifying a prior for $b$ and $\sigma$ implicitly defines a distribution for $R^2$. For instance, when $b_k$ and $\sigma$ have weakly informative priors \citep{GelmanCauchy}, as shown in \cite{aguilar_intuitive_2023}, the resulting prior for $R^2$ is highly concentrated near 1, even for moderate values of $K$. To illustrate the case of shrinkage priors, assume $X=I$ and use the well-known Horseshoe prior \citep{Horseshoe}, which is specified as:
%----------------------------
\begin{align}
    b_k | \lambda_k \sim \normal(0, \lambda_k^2) , \ \lambda_k \given \tau \sim C^+(0, \tau) , \tau | \sigma \sim C^+(0, \sigma),
\end{align}
%----------------------------
where $C^+(0, \tau)$ denotes a Half Cauchy distribution \citep{GelmanCauchy, PolsonHalfCauchy} with scale parameter $\tau$. We assume a fixed value for $\tau$ as a user-specified hyperparameter, rather than assigning it its own prior. As demonstrated by \cite{vanDP2016Conditions}, it is crucial for $\tau$ to reflect the proportion of signals in order to guarantee signal recovery.

\begin{figure}[t]
    \centering
    \includegraphics[width=0.90\linewidth]{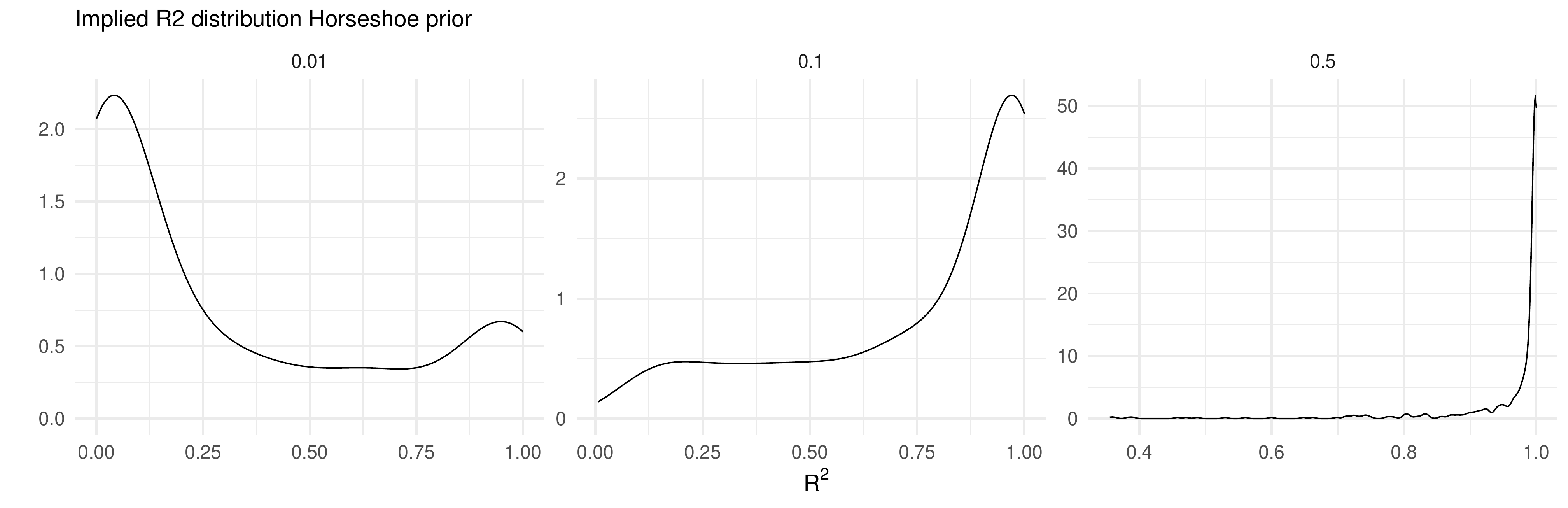}
    \label{fig:R2impliedHS}
    \caption{Shrinkage prior implied distribution for $\rsq$ when considering the Horseshoe prior for different values of $\tau$, which is interpreted as the a priori proportion of signals.}
\end{figure}

Figure \ref{fig:R2impliedHS} shows the implied distribution of $\rsq$ when using the Horseshoe prior with different values for $\tau$. Interpreting $\tau$ as the proportion of true signals, we observe that for very low $\tau$ values, the implied distribution of $\rsq$ tends to concentrate near zero. On the other hand, when a user believes that even a small proportion of the elements in $b$ are nonzero, i.e., $\tau \approx 0.1$, the distribution shifts toward one, indicating that the Horseshoe prior is focusing on identifying signals.

A significant issue arises when, despite having a low prior expectation for the proportion of nonzero coefficients, the model predicts high values of $\rsq$. This leads to overestimation of the importance and magnitude of potential signals, which can conflict with user intuition about how the number and strength of signals relate to the proportion of explained variance. Such an overestimation could mislead the user into thinking the model explains more variance than it realistically does, which could skew model interpretation and decision-making. Moreover, controlling the properties of the resulting distribution over $\rsq$, such as its mean and variance, becomes challenging, as they are implicitly determined by the prior and the data.
%----------------------------
%----------------------------
Given the total variance $\omega^2$ in the model, we can compute the \textit{proportion of explained variability} for each coefficient, defined as
%----------------------------
\begin{align}
    \phi_k \coloneqq \frac{\lambda_k^2}{\sum_{k=1}^K \lambda_k^2 }.
\end{align}
%----------------------------
Since $\phi_k \geq 0$ for all $k=1, \ldots, K$ and $\sum_{k=1}^K \phi_k = 1$, the vector $\phi \coloneqq (\phi_1, \ldots, \phi_K)$ lies in the $K-1$ dimensional simplex. Thus, the joint distribution of the $\lambda_k$ values induces a distribution on the simplex, represented stochastically by $\phi$. Understanding the implied distribution of $\phi$ provides insights into how the total variance $\omega^2$ is distributed among the coefficients and how they compete for it. This understanding can help determine how much of the total variance is allocated to each coefficient, which is key for assessing model fit and identifying influential predictors.

For example, if we assume $\lambda_k \sim \gammadist(\alpha_k, 1)$, we obtain the Dirichlet distribution $\phi \sim \dirichlet(\alpha)$, where $\alpha \coloneqq (\alpha_1, \ldots, \alpha_K)$ is the concentration vector \citep{OnTheDirichlet}. In this case, the correlations between the elements of $\phi$ would be negative and determined by the mean of the distribution. As with the implicit distribution for $\rsq$, specifying a prior on $\lambda$ does not necessarily give us direct control over the properties of the implied distribution for $\phi$, except in certain special cases, such as the one just mentioned.

The emphasis on implied quantities that are more intuitive and user-friendly has largely been overlooked in the shrinkage prior literature, with only a few notable exceptions \citep{ZhangCredibleRegions, r2d2zhang, aguilar_intuitive_2023}. This oversight is understandable, as statisticians have primarily concentrated on developing robust automatic procedures capable of tackling complex tasks. However, we believe that emphasizing the more intuitive perspective could not only enhance user comprehension and the practical applicability of well-established shrinkage priors but also inspire the development of new, innovative methodologies.

\subsection{Generalized Decomposition R2 priors}
\label{sec:genr2}

We have discussed how the specification of a shrinkage prior on the regression coefficients $b$ implies priors for interpretable quantities such as $\rsq$ and $\phi$. It is equally insightful to explore the reverse scenario, that is, establishing a prior over the proportion of explained variance $\rsq$ and the proportions of total variance $\phi$ to derive a prior over the regression coefficients $b$.  This concept has been explored before to define prior distributions in the context of additive regression models of varying complexity \citep{r2d2zhang, aguilar_intuitive_2023}. However, they always considered $\phi \sim \dirichlet(\cdot)$. While this choice is certainly a reasonable one, it raises the question of what happens when $\phi$ follows a distribution other than Dirichlet. 

Employing alternative distributions for $\phi$ allow us to probe how the proportions of total variance $\phi_k, k=1,...,K$ compete for the total variability $\omega^2$. Left unattended and without any proper control, this competition is naturally guided towards a certain degree of negative dependence structures, i.e., if $\phi_k$ increases then $\phi_j, j\neq k$ decreases since there is competition for a total quantity. Specifically, the Dirichlet distribution shows this phenomenon \citep{AitchisonMonograph, SimplexKruijer, OnTheDirichlet}. However, in reality, specific (groups of) coefficients may compete differently for the total variability than the Dirichlet would assume and allow for. Indeed, as we show in our simulations and real-world case studies, other prior choices may lead to more favourable results.

To this end, we introduce a prior distribution family that provides greater flexibility in capturing the intricate relationships among the proportions of explained variance. We call them \textit{Generalized Decomposition R2} (GDR2) priors. The main idea is schematically presented in Figure \ref{fig:sr2}. We begin by setting a prior over $\rsq$. According to Equation \eqref{eq::r2def}, there exists a a one to one relationship between $\rsq$ and the total variability $\omega^2$, immediately establishing a prior distribution on $\omega^2$. Subsequently, we select a distribution for the proportions of explained variance $\phi$ that can properly represent a desired dependence structure. Afterwards, we shape the variances of the regression coefficients $b_k$ by setting $\lambda_k^2= \phi_k \omega^2, i=1,...,K$. Finally, we let $b_k$ follow a normal distribution centered at zero with variance $\lambda_k^2$. 
%------------------------
\begin{figure}[t!]%
    \centering
    \begin{tikzpicture}
    % x node set with absolute coordinates

    % Define more fitting colors for Fantastic Mr. Fox
    \definecolor{foxorange}{rgb}{0.91, 0.592, 0.255}     % Warm orange for $R^2$
    \definecolor{foxyellow}{rgb}{1.0, 0.788, 0.027}    % Mustard yellow for $\omega^2$
    \definecolor{foxcream}{rgb}{0.949, 0.874, 0.816}     % Muted green for $\phi_k$
    \definecolor{foxmustard}{rgb}{0.765, 0.439, 0.129}    % Brown for $\lambda_k^2$
    \definecolor{foxred}{rgb}{0.776, 0.125, 0.153}         % Red for $b_k$

    \node[state, draw=foxorange, fill=foxorange!30] (r2) at (0,0) {$R^2$};

    \node[state, draw=foxyellow, fill=foxyellow!30] (t2) at (2,0) {$\omega^2$};

    \node[state, draw=foxcream, fill=foxcream!30] (phip) at (4,1) {$\phi_1$};
    \node[vmissing] (dots4) at (4,0) {};
    \node[state, draw=foxcream, fill=foxcream!30] (phip1) at (4,-1)  {$\phi_{K}$};

    \node[state, draw=foxmustard, fill=foxmustard!30] (lambdap) at (6,1) {$\lambda_1^2$};
    \node[vmissing] (dots6) at (6,0) {};
    \node[state, draw=foxmustard, fill=foxmustard!30] (lambdap1) at (6,-1) {$\lambda_{K}^2$};

    \node[state, draw=foxred, fill=foxred!30] (betap) at (8,1) {$b_1$};
    \node[vmissing] (dots8) at (8,0) {};
    \node[state, draw=foxred, fill=foxred!30] (betap1) at (8,-1) {$b_K$};

    \path (r2) edge (t2);
    \path (t2) edge  (phip);
    \path (t2) edge  (phip1);
    \path (t2) edge  (dots4);

    \path (phip) edge (lambdap) ;
    \path (phip1) edge (lambdap1) ;

    \path (lambdap) edge (betap);
    \path (lambdap1) edge (betap1);
    
\end{tikzpicture}
    \caption{Schematic of the GDR2 prior construction. First, a distribution is assigned to $R^2$, then its uncertainty is propagated to the regression terms via the simplex distribution of the proportions of explained variance. Finally, a distribution is used to allocate variance to each regression term.}
    \label{fig:sr2}
\end{figure}%
%------------------------
We opt for a Beta distribution to characterize $R^2$ and we write $R^2\sim \betadist (\cdot, \cdot )$. The Beta distribution, offering various parametrizations \citep{meanprecbeta}, is chosen for its flexibility. We specifically parametrize it in terms of the prior mean $\mu_{R^2}$ and prior "precision" $\varphi_{R^2}$ (also known as mean-sample size parameterization). This choice facilitates an intuitive integration of prior knowledge into the $\rsq$ prior. The hyperparameters $\mu_{R^2}, \varphi_{R^2}$ are interpretable expressions of domain knowledge that represent the existing relationship between the included covariates and the response variable. 

If $(a_1, a_2)$ are the canonical shape parameters of the Beta distribution, then $\mu_{R^2}= \frac{a_1}{a_1+a_2}$ and 
$\varphi_{R^2}= a_1+ a_2$. This selection implies that $\omega^2$ follows a Beta-Prime distribution, with parameters $\mu_{R^2}, \varphi_{R^2}$, which we denote as $\omega^2 \sim \betaprime(\mu_{R^2}, \varphi_{R^2})$. 
%----------------------------
\begin{figure}[b]
    \centering
    \includegraphics[width=1\linewidth]{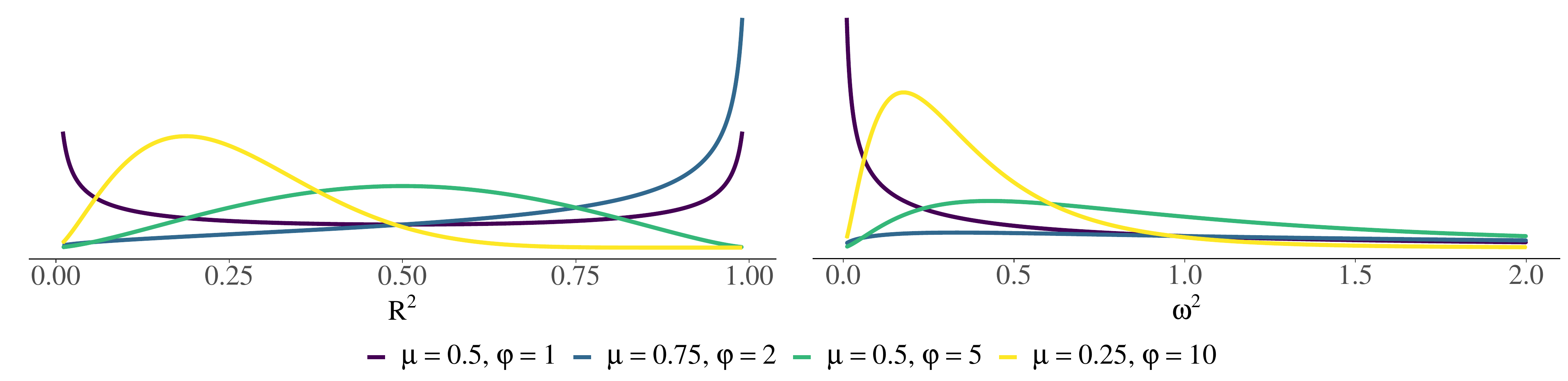}
    \vspace{-0.5cm}  
    \caption{Beta and Beta Prime densities for various values of the prior mean $\mu_{R^2}$ and precision $\varphi_{R^2}$. }
    \label{fig:betapriors}
\end{figure}
%----------------------------
Figure \ref{fig:betapriors} illustrates the flexibility that the Beta distribution can offer in expressing prior knowledge about $R^2$. The figure also depicts the corresponding Beta Prime prior for $\omega^2$. For instance, for values of $(\mu_{R^2},\varphi_{R^2})=(0.5,1)$, we obtain a bathtub-shaped prior on $R^2$, concentrating the majority of the mass near the extremes $R^2=0$ and $R^2=1$. This prior behavior signals the user's expectation of a model containing either a substantial amount of either noise or signal, respectively. 

Moving forward, the \textit{(probability) simplex} of dimension $K-1$ ($K \in \mathbb{N}$), is defined as $\mcs^{K} \coloneqq  \{ x \in \mathbb{R}^K  | \sum_{k=1}^K x_k=1, x_k \geq 0 ,k=1,...,K \}$. Let $p(\phi \, | \,\nu)$ represent the chosen distribution for $\phi$ on the simplex $\mathcal{S}^k$, where $\nu$ represents specific hyperparameters related to the chosen distribution. For example, in the case of the Dirichlet distribution $\nu = \{\alpha$\}. Although the Dirichlet distribution aids in providing analytical results, interpretation, computational efficiency, and alignment with common user knowledge, it exhibits undesirable properties in practice, which we discuss further in Section \ref{sec:simplexdists}.

The next step is to specify a distribution for the coefficients that captures the attributed variance for each regression coefficient $b_k$, given by $\lambda_k^2 = \phi_k \omega^2 \sigma^2$ for $k = 1, \dots, K$. To do so, we define $b_k | \sigma, \lambda_k \sim \normal(0, \sigma \lambda_k)$, which integrates our prior into the GL shrinkage prior framework, typically modeled as a scale mixture of normals \citep{mwestScaleNormals}. In contrast, \cite{DirichletLaplace} and \cite{r2d2zhang} have used double exponential distributions for $b_k$ to express the attributed variances, although the former did not account for the perspective of variance proportions. Additionally, \cite{aguilar_intuitive_2023} adopted normal distributions to model the attributed variances, offering a different approach to the distribution of the coefficients.

If model \eqref{eqn::linreg} includes an intercept $b_0$, the prior specification is simplified by setting a prior on the centered intercept $\tilde{b}_0$, which is implied when $\mathbb{E}(x_i) = 0$. The original intercept $b_0$ can then be recovered after model fitting using a simple linear transformation \citep{brmsJSS, rstanarm}. Common choices for priors on $b_0$ include a normal prior with mean $\mathbb{E}(y)$ and a user-chosen scale, which depends on the scale of $y$, or Jeffrey’s prior, which is improper and flat in this case \citep{Jeffreys}. Next, we specify a prior for the residual variance $\sigma^2$ (or equivalently the residual standard deviation $\sigma$). Following the recommendations of \cite{GelmanHalfStudentt}, we set a Half Student-$t$ prior on $\sigma$ with $\nu$ degrees of freedom and scale $\eta$. Consistent with \cite{brmsJSS} and \cite{aguilar_intuitive_2023}, we set $\eta \approx \text{sd}(y)$, as both the prior’s expected mean and variance are proportional to $\eta$. Together, the full GDR2 model is given by
%----------------------
\begin{equation}
\label{gdr2model}
    \begin{aligned}
    y_n & \sim \normal(\mu_n, \sigma^2), \quad \mu_n = b_0 + \sum_{k=1}^K x_{nk} b_k, \quad n = 1, \dots, N, \\
    b_0 & \sim p(\cdot), \quad b_k \sim \normal(0, \sigma^2 \phi_k \omega^2), \quad k = 1, \dots, K, \\
    \omega^2 & = \frac{R^2}{1 - R^2}, \quad \phi \sim p(\nu_{\phi}), \quad R^2 \sim \betadist(\mu_{R^2}, \varphi_{R^2}), \quad \sigma \sim p(.).
    \end{aligned}
\end{equation}
%----------------------
While, in this paper, we focus on the linear regression case, the additive structure of the prior readily allows its extension to linear multilevel models or other kinds of additive models if desired. 

\subsection{Distributions on the Simplex}
\label{sec:simplexdists}
In this section, we explore potential distributions for $\phi$ defined on the simplex. While extensions of the Dirichlet distribution are often considered, we do not pursue them here, as they share similar limitations \citep{ConnorMossimanGenDirichlet, simplex_nielsen, simplex_ongaro,chow2022schlomilchintegralsprobabilitydistributions}. Instead, we draw on methods from Compositional Data Analysis (CDA) \citep{AitchisonMonograph, CompositionsR, Aitchison40years}, Categorical Data Analysis (CatDA)\citep{CATDAagresti_bayesian_2005} and Correlated Topic Modeling (CTM) \citep{CorTM, ScalableCTMLN}, where the logistic normal distribution (and its variants) is the preferred approach for analyzing simplicial variables. 

\subsubsection{The Dirichlet distribution}

The Dirichlet distribution is widely employed for modeling data existing in the simplex \citep{Dirichlet_history, OnTheDirichlet}. A vector $\phi=(\phi_1,...,\phi_K)' \in \mcs^{K} $ follows a Dirichlet distribution with concentration vector $\alpha \coloneqq (\alpha_1, ..., \alpha_K )', \alpha_k >0 , k=1,..., K$ if its density has the following form:
%-------------
\begin{align}
\label{eq:dirdist}
p(\phi | \alpha)&= \frac{\Gamma(\alpha_+ )}{ \prod_{k=1}^K \Gamma(\alpha_k) } \prod_{k=1}^K \phi_k^{\alpha_k-1} \mathbbm{1}(\phi \in \mcs^{k}),
\end{align}
%-------------
where $\alpha_+= \sum_{k=1}^K \alpha_k$. The mean and covariance of the distribution are respectively expressed as: $\mbe(\phi_k) = \frac{ \alpha_k }{\alpha_+} =: \tilde{\alpha}_k$ and $\cov(\phi_k, \phi_j)= \frac{ \delta_{kj} \tilde{\alpha}_k - \tilde{\alpha}_k \tilde{\alpha}_j }{1+\alpha_+ }$,
where $\delta_{kj}$ is the Kronecker delta function. If $k \neq j$, then $\cov(\phi_k, \phi_j) = \frac{ - \tilde{\alpha}_k \tilde{\alpha}_j }{\alpha_+ +1} <0$ for all $k,j$. Once the mean vector $\tilde{\alpha}$ is chosen, only the scalar $\alpha_+$ remains to express the entire variance-covariance structure. In particular, the covariance function is always proportional to the product of the corresponding means. 

The Dirichlet distribution gained prominence during the conjugate prior era, as it serves as the conjugate prior for the multinomial likelihood \citep{simplex_ongaro, OnTheDirichlet, wang2024pochhammerpriorssparsecount}. Its interpretability is another advantage, the hyperparameters $\alpha_k$ can easily be interpreted in relation to the distribution's behavior. In particular, the quantity $\tilde{\alpha}_k = \mbe(\phi_k)$ represents the relative a priori importance of the $k$th element $\phi_k$. To illustrate, consider $\alpha=(1,...,1)'$, which gives a distribution that is flat over all possible simplexes, leading to a suitable choice in the absence of additional prior knowledge. Generally, setting $\alpha=(a_\pi,...,a_\pi)'$ with a concentration parameter $a_\pi > 0$ (i.e., a \textit{symmetric Dirichlet distribution}) drastically reduces the number of hyperparameters to specify and allows us to globally control the shape of the Dirichlet distribution with a single value. A symmetric Dirichlet distribution is useful when there is no prior preference that favors one component over another. Values $a_\pi<1$ concentrate mass on the edges of the simplex, while $a_\pi>1$ results in concentration around the center of the simplex. The user can also specify asymmetric Dirichlet distributions by choosing different values for the individual $\alpha_k$. This represents different a priori expected importance of the corresponding components.

A major limitation of the Dirichlet distribution is its inherent lack of flexibility in modeling dependencies among the elements of $\phi$. Its covariance structure is determined entirely by the mean, restricting the ability to specify custom dependency structures for simplex data. Moreover, the sum-to-one constraint induces negative correlations between components, making the Dirichlet unsuitable for scenarios where components may exhibit positive correlations \citep{ConnorMossimanGenDirichlet, GenDirWONG1998}. Even many complex negative correlation patterns cannot be accurately captured, further highlighting the limitations of the Dirichlet distribution in accommodating diverse dependency structures.

From a technical perspective, the Dirichlet distribution imposes additional constraints that limit its applicability. It requires properties such as closure under marginalization and conditioning, as well as complete subcompositional independence. This last property implies that all renormalized subsets of $\phi$ are independent of each other \citep{AitchisonMonograph}. Subcompositional independence has been widely criticized, as it imposes $1/2K(K-3)$ constraints on the covariance structure, rendering the Dirichlet implausible for most real-world applications \citep{logisticnormal, AitchisonMonograph, simplex_ongaro}. These limitations make the Dirichlet unsuitable for modeling complex or realistic dependency structures, necessitating more flexible alternatives like the Logistic Normal distribution, which we present below.

\subsubsection{The Logistic Normal Distribution}

The logistic normal distribution, proposed by \cite{logisticnormal} offers a solution to overcoming limited correlation structures. The core concept involves mapping a multivariate normal random variable defined from $\mathbb{R}^{K}$ into the $K-1$ dimensional simplex $\mathcal{S}^{K}$ by the use of log ratio transformations, as described below. This approach enables researchers to make use of well known established techniques for multivariate analysis in the unbounded real space and then seamlessly transforming the results into the simplex \citep{AitchisonMonograph, PawlowskyCompositionalDA}.
   
Let $\phi \in \mathcal{S}^K$, the \textit{additive log-ratio} (alr) transformation, relative to the $K$th component, is defined as: \citep{logisticnormal} $\eta= \alr(\phi) \coloneqq \left(  \ln \left( \frac{\phi_{1}}{\phi_{K}} \right), ...,  \ln \left( \frac{\phi_{K-1}}{\phi_{K}} \right)  \right) \in \mathbb{R}^{K-1}$.

This transformation establishes a one-to-one correspondence between the elements of the simplex and the log-ratio vectors $\ln \left( \frac{\phi_i}{\phi_K} \right)$, allowing any statement about the components of the simplex to be equivalently expressed in terms of these log-ratios \citep{logisticnormal, AitchisonMonograph}. By mapping compositional data into the unconstrained real space $\mathbb{R}^{K-1}$, the difficulties associated with working in a constrained space are effectively removed.

The alr transformation is inherently asymmetric since the log-ratios require a reference component. The last component $\phi_K$ is usually chosen as reference, but any component can be selected based on interpretability or practical considerations \citep{AitchisonMonograph}. Guidelines for choosing the reference component have been proposed \citep{Aitchison40years}, but in general, practitioners of compositional data analysis (CDA) and correlated topic models (CTM) recognize that the choice of reference has minimal impact on results or inference \citep{PawlowskyCompositionalDA, CompositionsR, Aitchison40years}. 
%Moreover, \cite{AitchisonMonograph} proves that statistical inference using the alr transformation is independent of both the reference component and the ordering of $\phi$. 

The inverse alr transformation is defined by the\textit{ standard logistic} (\textit{softmax}) transformation with a fill-up term for the reference component: $\phi_k= \frac{ e^{\eta_k}}{1+\sum_{j=1}^{K-1} e^{\eta_j} }, k=1,..., K-1, ,  \phi_K= 1-\sum_{k=1}^{K-1}\phi_k.$ We say that $\phi \in \mathcal{S}^K$ follows an \textit{additive logistic normal} (ALN) distribution if $\eta = \alr(\phi)$ follows a multivariate normal distribution in $\mathbb{R}^{K-1}$. The density of the ALN distribution is given by \citep{AitchisonMonograph}:
%--------------------------%
\begin{equation}
    \begin{aligned}
\label{eqn:aln}
 p(\phi | \mu, \Sigma)| &= | 2 \pi \Sigma|^{-1/2} \left( \prod_{k=1}^K \phi_k \right)^{-1} \exp\left \lbrace -\frac{1}{2} \left[  \alr{(\phi)} - \mu \right]' \Sigma^{-1}  \left[ \alr{(\phi)} - \mu \right] \right \rbrace.
    \end{aligned}
\end{equation}
%--------------------------%
We denote this as $\phi \sim \text{ALR}(\mu, \Sigma)$, where $\mu$ and $\Sigma$ depend on the chosen reference component. 

A symmetric representation for $\phi$, independent of a reference component, can be obtained by starting from the unconstrained space and mapping into the simplex. Specifically, let $\eta \sim \normal_K(\mu, \Sigma)$ and apply the \textit{symmetric logistic transformation}: 
%---------------------------------------
\begin{align}
\label{eq:logistic_map}
\phi_k = \frac{e^{\eta_k}}{\sum_{j=1}^K e^{\eta_j}} , \ k=1, \ldots, K.
\end{align}
%---------------------------------------
In this case, $\phi$ is said to follow a (symmetric) \textit{Logistic Normal} (LN) distribution with parameters $\mu, \Sigma$, denoted as $\logisticnormal(\mu, \Sigma)$. For identifiability, either the constraint $\sum_{k=1}^K \eta_k = 0$ or $\eta_K = 0$ can be imposed, the latter being equivalent to the ALR \citep{bishop2006pattern, CorTM, Goodfellow}. If we consider the sum-to-zero constraint, then $\eta_k = \ln \left( \frac{\phi_k}{g(\phi)} \right)$, where $g(\phi)$ denotes the geometric mean of $\phi$. 

Importantly, $\Sigma = (\sigma_{ij}), i, j = 1, \ldots, K$, represents the covariance structure of the log-ratios, not of the raw proportions $\cov(\phi_i, \phi_j)$. If one chooses to model $\phi$ using either the ALR or LN distribution, transitioning between them is entirely feasible via a straightforward linear relationships that we show in the Appendix \ref{appendix}. This flexibility allows researchers to select the representation most suited to the practical requirements of their analysis.

While moments of all non-negative orders, $\mbe_\phi\left( \prod_{k=1}^K \phi_k^{m_k} \right), \, m_k \geq 0, \, k=1, \ldots, K$, exist, their expressions are not reducible to simple analytic forms \citep{logisticnormal, MomentsLogitNormal_2, Momentslogitnormal}. Nonetheless, when analyzing quantities in the simplex, it is customary to study ratios or logarithms of ratios of simplex components rather than raw proportions \citep{AitchisonMonograph, CompositionsR, PawlowskyCompositionalDA, CreusMart2021ADA}. The primary motivation for this is the so-called \textit{coherence} requirement in compositional data analysis, which ensures that the relationships between components remain consistent and interpretable, whether considering the full composition or its subcompositions \cite{Aitchison40years}. Coherence is especially important since the constant-sum constraint in compositional data implies that changes in one component inherently affect all others.

Let \( \sigma_{ik, jl} = \cov\left( \ln\left( \phi_i / \phi_k \right), \ln\left( \phi_j / \phi_l \right) \right) \) represent the covariance of log-ratio quantities, then the following key relationships hold:
%---------------------------------
\begin{align}
\label{eq:logitnorm_mucov}
     \mbe\left \lbrace  \ln \left( \phi_j/\phi_k \right)  \right \rbrace &= \mu_j - \mu_k , \ \ 
      \sigma_{ik, jl} =   \sigma_{ij}+\sigma_{kl}-\sigma_{il}-\sigma_{jk}.
     % \mbe \left( \phi_j/ \phi_k \right) &= \exp \left \lbrace \mu_j - \mu_k + 1/2\left( \sigma_{jj}- 2\sigma_{jk}+ \sigma_{kk} \right) \right \rbrace \\
    % \cov\left( \phi_j / \phi_k, \phi_l / \phi_m  \right) &= \mbe \left( \phi_j/ \phi_k \right) \mbe \left( \phi_l/ \phi_m \right) \exp \left\lbrace  (\sigma_{jl}+\sigma_{km}-\sigma_{jm}-\sigma_{kl}) -1  \right\rbrace
\end{align}
%--------------------------
Thus, the mean vector  $\mu$  and covariance matrix $\Sigma$ fully determine the expectations and pairwise covariances of log-ratio quantities. This implies that no recalculations are necessary when transitioning between different log-ratio representations, as all relevant information is encoded in $\mu$ and $\Sigma$. Moreover, one can derive $\mu$ and $\Sigma$ directly from knowledge of any pair of log-ratio quantities, and vice versa \citep{AitchisonMonograph}. We explore these parameters in greater detail in Section~\ref{sec:hyperparam_spec}, where we incorporate the LN distribution as a prior for  $\phi$  within the GDR2 framework. 

\subsection{Hyperparameter Specification in GDR2 priors}
\label{sec:hyperparam_spec}

As explained in Section \ref{sec:genr2}, the Beta prior for $R^2$ can be parametrized in terms of the prior mean $\mu_{R^2}$ and precision $\varphi_{R^2}$. In cases where the user has limited prior knowledge, a flat distribution over the unit interval can be specified by setting $(\mu_{R^2}, \varphi_{R^2}) = (0.5, 2)$. Alternatively, a bathtub-shaped distribution can be chosen with $(\mu_{R^2}, \varphi_{R^2}) = (0.5, 1)$, which places more mass on extreme values of $R^2$. This reflects a prior belief that either there is no relationship between the response and the covariates, or that the covariates fully explain the variability in the response.  An application of the law of Total Variance shows that $\var(b) = \mathbb{E}[\var(b \mid \phi, \omega^2)] + \var[\mathbb{E}(b | \phi, \omega^2)] = \mathbb{E}(\omega^2) \mathbb{E}(\text{cov}(\phi))$. Thus, the hyperparameters selected for $\omega^2$ can control the tail behavior of $b_k$, potentially inducing infinite variance and heavy tails in its marginal distribution. Such a configuration enhances sensitivity to detecting signals and is particularly appropriate when heavy tails are desired in the regression coefficients.

As discussed in Section $\ref{sec:simplexdists}$, we let $\phi$ follow either a Dirichlet or an LN  distribution. When setting $\phi\sim \dirichlet(\alpha)$ the GDR2 prior is equivalent to the R2D2 prior \citep{r2d2zhang} for non-hierarchical models, except that $b_k$ follows a normal distribution rather than a Double Exponential. The former scenario has been explored by \cite{aguilar_intuitive_2023}, who show that if $\phi \sim \dirichlet(\alpha)$ with $\alpha=(a_\pi,..., a_\pi)$, then $a_\pi$ determines the behavior of the marginal distribution of $b_k$ near the origin. Thus, $a_\pi$ can be theoretically chosen to tailor the properties of the marginal distribution of $b_k$. As an alternative, the elements of $\alpha$ can be elicited from subject matter experts to represent the a priori relative importance of covariates -- although we consider it very hard elicit such knowledge for high-dimensional problems. 

If $\phi \sim \logisticnormal(\mu, \Sigma)$, the prior specification involves defining the mean vector $\mu$ and covariance matrix $\Sigma$ for the log-ratios. Setting $\mu_k = 0 \ \forall k$ (Equation \eqref{eq:logitnorm_mucov}) encodes the belief that all proportions $\phi_k$ are equally weighted, expressing no preference for any specific component. To assign equal weights to a subset of proportions, $\mu_i = c$ can be specified for $i \in I \subset \{1, \ldots, K-1\}$. If the proportions can be grouped into $G$ mutually exclusive sets $I_g \subset \{1, \ldots, K-1\}$, group-specific weights $c_g$ may reflect their relative importance, enabling the inclusion of prior hierarchical or structural information.

To better understand the effect of assigning values to $\mu$, consider the symmetric logistic mapping in Equation \eqref{eq:logistic_map} and assume $\Sigma = \sigma_\phi^2 I$. Setting $\mu_k = c \ \forall k$ is equivalent to $\mu_k = 0$, as the logistic mapping is shift-invariant \citep{bishop2006pattern}. This choice reflects a lack of prior knowledge about relative importance and by symmetry it can be shown that $\mathbb{E}[\phi_k] = 1/K$. When $\mu_k = c_k$, proportions will differ depending on the sign of $c_k$; components with $ c_k > 0$ will, on average, be larger, while those with $c_k <0$ smaller. To see this, let $R_k = e^{\eta_k}$ and $S = \sum_j e^{\eta_j}$. Using a first order Taylor expansion around the mean, we can approximate \citep{stuart2009kendall}:
%-------------------
\begin{align}
\label{eq:logittaylor}
\mathbb{E}[\phi_k] = \mathbb{E}\left(\frac{e^{\eta_k}}{\sum_j e^{\eta_j}}\right) \approx \frac{\mathbb{E}[R_k]}{\mathbb{E}[S]} = \frac{  e^{ c_k +{\frac {\sigma ^{2}}{2}}} }{ \sum_j e^{ c_j +{\frac {\sigma ^{2}}{2}}} } = \frac{ e^{c_k}  }{ \sum_j e^{ c_j  }},
\end{align}
%--------------------
where we compute the expectations by noting that $R_k$ follows a log normal distribution. This argument extends to a diagonal covariance matrix $\Sigma$ with different scales $\sigma_j$; whose values interact with $c_j$ to redistribute mass in favor or against $\phi_k$.

The relationship becomes more complex for a general  $\Sigma$ and $\mu$. While a Taylor expansion provides an approximation\footnote{see Appendix \ref{appendix}}, it relies on the term \text{Cov}$(R_k, S)$, which lacks a closed-form solution, requiring Monte Carlo methods for estimation. While this offers insights into the interaction of key quantities that determine the marginal expectation, its added complexity may outweigh its utility. Our experiments in Section \ref{sec:experiments} show that, setting $\mu_k = 0$ is a robust default choice in the absence of prior knowledge, balancing simplicity and performance.
%----------------------------
\begin{figure}[t]
    \centering
    \includegraphics[width=1\linewidth]{figs/prior_plots/ImpliedNorm.pdf}
    \vspace{-0.5cm}  
    \caption{Implied distribution on the norm of a Multivariate Normal as the dimension $K$ increases. The red line shows the value  $\sqrt{K} \sigma_\phi$, which is the point of concentration, where $K$ is the number of covariates and $\sigma_\phi = 1$. }
    \label{fig:ImpliedNorm}
\end{figure}
%----------------------------
The specification of $\Sigma$ can be understood by analyzing the implied prior on the norm of $\eta$ (the log ratios) and how this translates to the simplex. We first consider the case of a diagonal $\mu = 0, \Sigma = \sigma_\phi^2 I$. Let $\eta \sim N(0, \sigma_\phi^2 I)$, since $\eta$ is a sub-gaussian variable with independent entries \citep{giraud2014HD, vershynin2018high}, $\|\eta\|$ concentrates around $\sqrt{K}\sigma_\phi$. Specifically, the following two-sided tail bound holds:  $\mathbb{P} \left( \left| \frac{ \|\eta\| }{ \sqrt{K}\sigma_\phi } - 1 \right| \geq t \right) \leq 2\exp\left(-\frac{K \sigma_\phi^2  t^2}{ 2 C}\right)$ for a constant $C>0$. We illustrate this behavior in Figure \ref{fig:ImpliedNorm}.

The user-specified $\sigma_\phi$ defines bounds for the norm of the logits, setting a “budget” that can be distributed across them. For $\sigma_\phi \to 0$, the logits concentrate near zero, resulting in $\phi_k \to 1/K$. Conversely, as $\sigma_\phi$ increases, the logits spread further, increasing the quota for $\|\eta\|$. If $\mu \neq 0$, nonsparse elements will be determined by the locations in which $\mu_k \neq 0$, especially if the budget is low. 

The choice of $\sigma_\phi$ balances the variability in the simplex as $K$ grows. Setting $\sigma_\phi = \sqrt{ \gamma/K}, \gamma > 0$ ensures that $\|\eta\|^2 \approx \gamma$, maintaining a consistent magnitude for the logits regardless of $K$. Alternatively, choosing $\sigma_\phi = \sqrt{\gamma}$ (independent of K) allows the logits’ norm to grow proportionally to $\sqrt{K}$, introducing greater variability. This variability interacts with $\mu$ to influence how mass is distributed across $\phi_k$. For scenarios where sparsity is critical, setting $\sigma_\phi = \sqrt{\log(K)/K}$ slows the growth of the logits’ norm, preventing excessively large values and reducing variability. This encourages sparse distributions, with mass concentrated at locations dictated by $\mu$. Such a setup is particularly effective in high-dimensional settings. Users must carefully consider the interaction between $\mu$ and $\sigma_\phi$. Lower values of $\sigma_\phi$ concentrate the mass of $\phi$ at positions where $\mu$ has higher values, promoting focused distributions. In contrast, higher values of $\sigma_\phi$ introduce variability, allowing for more diffuse distributions across the simplex.

When $\Sigma$ incorporates non-zero correlations, deriving a general inequality for the concentration of measure of $\|\eta\|$ becomes significantly more complex. We do not explore this case further, since we consider the intuition provided by $\sigma_\phi$ sufficient for practical specification. However, if users have prior knowledge about how the variables are correlated, they can include this information via a correlation matrix $\Omega_\phi$, using the decomposition $\Sigma = \text{D}_{\phi} \Omega_\phi \text{D}_{\phi}$, where $\text{D}_{\phi}$ is a diagonal matrix containing the scales, and $\Omega_\phi$ is a correlation matrix. For example, if the coefficients are ordered and proportions are expected to be similar for adjacent variables, one might use an autoregressive correlation structure. Alternatively, for cases where groups of variables exhibit distinct interdependencies, a block correlation matrix may be appropriate. This flexibility allows users to encode meaningful structural relationships directly into the prior.

\subsubsection{Prior matching}

An alternative approach to specifying the hyperparameters of the LN distribution is to draw upon theoretical insights and practical experience from the $\phi \sim \dirichlet(\alpha)$ case. Specifically, we propose using the Dirichlet distribution as an initial guide to determine the hyperparameters of the LN distribution. Our method involves defining a Dirichlet distribution that captures the desired theoretical and practical characteristics of the prior. Then, we minimize a divergence measure between the Dirichlet and LN distributions. This automated process, which we term \textit{prior matching}, provides a systematic way to derive the prior mean and covariance matrix for the LN distribution.

Prior matching simplifies the complex task of specifying LN hyperparameters by using the more analytically tractable Dirichlet distribution as a reference. While the LN distribution offers increased flexibility compared to the Dirichlet, its analytical properties can be challenging to study, particularly in the context of shrinkage priors. Prior matching mitigates these challenges, automating the derivation of the LN’s prior mean and covariance matrix for improved efficiency and user convenience.

Exact matching is not the ultimate goal, as our objective is to develop priors that outperform the Dirichlet. Prior matching serves as a starting point, enabling users to leverage the unique flexibility of the LN distribution. For instance, the LN distribution addresses a critical limitation of the symmetric Dirichlet distribution: sampling difficulties and numerical instability as $a_\pi \to 0$.

As divergence metric for prior matching, we propose to use the Kullback Leibler (KL) divergence \citep{KLdiv}. The KL divergence between two probability distributions $f,g$ is defined as  $\KL{f}{g} \coloneqq \mbe_f \left[  \ln\left( \frac{f(x)}{g(x)} \right)  \right] = \int_{-\infty}^\infty \ln \left( \frac{  f(x) }{ g(x)}\right) f(x)  dx.$ To find the LN distribution that best approximates a given Dirichlet distribution, we minimize the KL divergence between them. Letting $f \sim \dirichlet(\alpha)$ with $\alpha$ fixed and $g \sim \LN(\mu, \Sigma)$, the optimal LN parameters are obtained by solving: $ \mu^*, \Sigma^* = \argmin_{\mu, \Sigma}  \KL{f}{g}.$ The closed form solution is given by 
%----------------------------
\begin{align}
\label{eq:hyperparamskl}
    \mu_k^*&= \delta(\alpha_k)- \delta(\alpha_K), \sigma_{kk}^*= \varepsilon(\alpha_k)+\varepsilon(\alpha_{K}), \ k =1,..., K,\ \  \sigma_{kj}^*= \varepsilon(\alpha_K)  \ (k \neq j),
\end{align}
%----------------------------
where $\delta(x)= \frac{\Gamma'(x)}{\Gamma(x)}$ and $\varepsilon(x)=\delta'(x)$ are the digamma and trigamma functions respectively \citep{logisticnormal, AitchisonMonograph, Zwillinger}. Given their analytical form, these quantities can be computed instantly, adding no computational overhead to the procedure. While other divergence measures could be considered, the closed-form solution for the KL divergence offers a clear advantage in terms of both practical applicability and computational efficiency.

To exemplify KL matching, consider a symmetric Dirichlet distribution with a single hyperparameter $a_\pi$.  The value of $a_\pi$ controls the amount of shrinkage induced by the R2D2 prior \citep{aguilar_intuitive_2023}. When $a_\pi \leq 1/2$ the prior will be unbounded near the origin, leading to strong shrinkage towards zero. Users who wish to replicate this shrinkage behavior in the LN distribution can propose a value of $a_\pi \leq 1/2$ and use Equation \eqref{eq:hyperparamskl} to calculate the corresponding LN hyperparameters: $\mu_k=0, \sigma_{kk}=2 \varepsilon(a_\pi)$, and $\sigma_{kj}= \varepsilon(a_\pi)$. As $a_\pi$ increases, the KL divergence between the Dirichlet and LN distributions decreases \citep{AitchisonMonograph}. When $a_\pi \to 0$, the KL divergence increases significantly since the Dirichlet allocates mass along the simplex edges, a challenge for the LN distribution to approximate, given that it is defined within the simplex’s interior.
%----------------------------
\begin{figure}[t]
    \centering
    \includegraphics[width=1\linewidth]{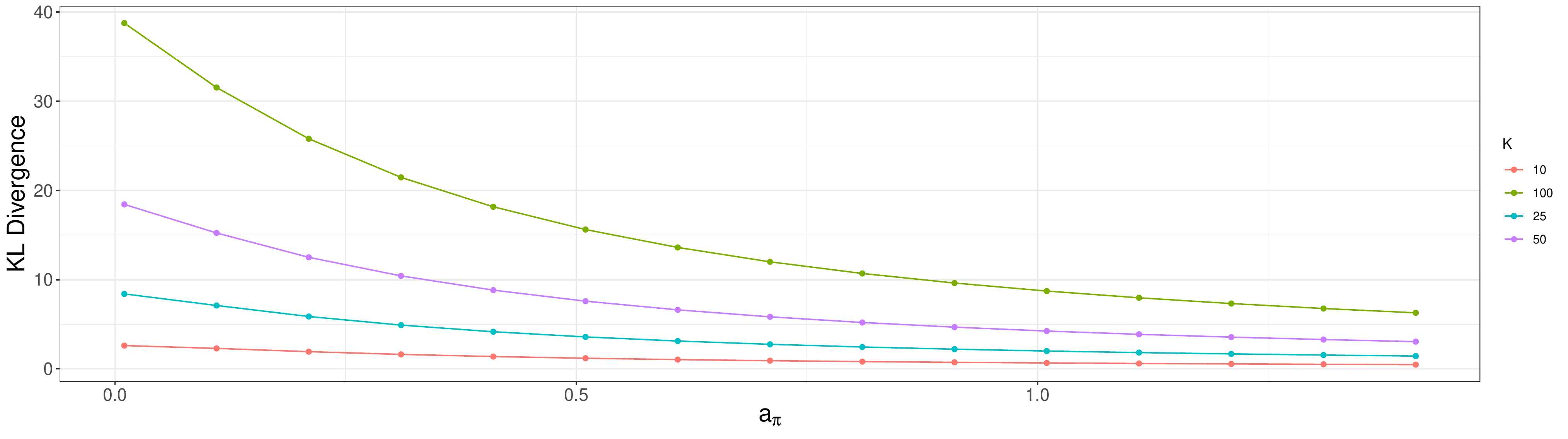}
    \vspace{-0.5cm}  
    \caption{ KL divergence between a symmetric Dirichlet distribution (parameter $a_\pi$) and the closest logistic normal distribution as $a_\pi$ and the number of covariates ($K$) vary. KL divergence decreases monotonically with $a_\pi$.}
    \label{fig:KLDirLN}
\end{figure}
%----------------------------
This is illustrated in Figure \ref{fig:KLDirLN} for varying values of $a_\pi$ and different dimensions $K$. 

A straightforward modification to KL matching involves discarding the correlations in the derived $\Sigma^*$, using only its diagonal elements to create a diagonal covariance matrix $\widetilde{\Sigma}^*$. In this case the scales $ \sigma_{kk}^2= 2 \varepsilon(a_\pi), \forall k$ and $\text{cor}(\eta_i, \eta_j) = 1/2, i \neq j$. In high dimensional settings, it is unrealistic to expect such correlations as a default, as we typically anticipate only a small number of signals \citep{piranha1}. Extracting only the scales, while discarding the correlations, is sensible since this type of correlations may be rendered as noise or irrealistic in high dimensions. The deviation between $\widetilde{\Sigma}^*$ and $\Sigma^*$ can be quantified by the KL divergence in log-ratio space. In the case where $\Sigma^*$ is obtained through KL matching with a symmetric Dirichlet distribution with hyperparameter $a_\pi$, this divergence has a simple: $\frac{1}{2} \ln \left( \frac{2^{K-1}}{K} \right)$, which is independent of $a_\pi$. 

We conclude this section by noting that discussions on selecting hyperparameter values for the Logistic Normal distribution as a prior are limited, and practical guidance is often scarce. Researchers often prefer to set priors over the hyperparameters  \citep{CorTM, CTMmimno2008gibbs, CTMijcai2017p588}, either by using priors of the form $\mu \mid \Sigma  \sim \normal(\mu_0, \gamma \Sigma)$ with $\Sigma \mid W \sim \text{InverseWishart}(\delta, W^{-1})$ \citep{ScalableCTMLN}, or by employing a Variational Approximation to the posterior distribution, where the variational family assumes a diagonal structure for $\Sigma$ \citep{CorTM, BleiVI}.  However, these methods often lack sufficient discussion on the choice of hyperpriors for the hyperparameters. In this work, we opted not to explore these approaches due to their added computational complexity, comparable performance with fixed hyperparameters, and the limited understanding of how such specifications interact with shrinkage priors in fixed-sample settings.

\section{Experiments and case studies}
\label{sec:experiments}

We conducted simulation studies to assess the performance of our generalized R2 priors under various conditions commonly encountered in practical scenarios. These investigations provide insights into how the priors behaves with finite samples and explores different possibilities for the total variance decomposition. Additionally, we present case studies utilizing real-world datasets that are commonly used as benchmarks for shrinkage priors.

The shrinkage priors considered in our simulations include the Beta Prime, Dirichlet-Laplace, Horseshoe, and the proposed GDR2 model, which incorporates Dirichlet and Logistic Normal decompositions. While these priors provide a useful context for comparison, our primary focus is on the GDR2 model, particularly its variance decomposition mechanisms. Results for the other priors, evaluated using their default settings, are detailed in  \cite{aguilar2024suppl}. Notably, the GDR2 model demonstrates comparable performance to these well-established priors and even outperforms them in specific scenarios. This highlights the robustness of GDR2 and underscores the flexibility and potential of decomposition-based approaches.

All models were implemented in the probabilistic programming language Stan \citep{StanJSS, stan2022}, which provides an extended implementation of the No-U-Turn Sampler (NUTS) \citep{hoffman2011nouturn}, an adaptive form of Hamiltonian Monte Carlo (HMC) \citep{handbookmcmc}. The Stan code for the GDR2 prior is included in Appendix \ref{appendix}, and all associated data and code can be found in \myosfresults.

\subsection{Simulations}
%-----------------------------
\subsubsection{Generative Models}

We use model \eqref{eqn::linreg} as a generative model in our simulations, adapting it to accommodate various forms of data encountered in practice, including diverse levels of sparsity and different dependency structures among the covariates. The design matrix $X$ was sampled from a multivariate normal distribution with mean 0 and covariance matrix $\Sigma_x$ derived from an AR(1) correlation structure with $\rho_x \in \{0, 0.8\}$. The intercept $b_0$ was drawn from a normal distribution with mean zero and variance $\sigma_I^2 = 4$. We fixed the sample size at $N = 100$, and varied the number of regression coefficients $K \in \{50, 150, 750\}$ to represent both low-dimensional ($K < N$) and high-dimensional ($K \gg N$) scenarios. The residual standard deviation $\sigma$ was adjusted using Equation \eqref{eq::r2def} to maintain the true explained variance $R^2_0 \in \{0.25, 0.6\}$.

The regression coefficients $b_k$, were generated in two ways: 1) \textit{Simulated Coefficients:} Coefficients were sampled from a normal distribution with zero mean and covariance $\Sigma_b$. Two forms of $\Sigma_b$ were considered: (A) diagonal with $\sigma_b^2 = 9$, and (B) AR(1) with autocorrelation $\rho_b = 0.8$ and diagonal elements $\sigma_b^2 = 9$. Sparsity was induced by setting each coefficient to zero with probability $v = 0.75$, reflecting real-world scenarios where coefficients arise from random processes \citep{DirichletLaplace, griffin2010inference, Griffinsomepriors}. 2) \textit{Fixed Coefficients}: This setup is akin to examples encountered in the shrinkage prior literature \citep{Horseshoe, griffin2010inference, DirichletLaplace, r2d2zhang}. We place concentrated signals $b^* \in \{3, 7\}$ in specific locations of the coefficient vector $b$. The first 5 and the last 5 elements are set to $b^*$, while all others are set to 0.

We have specified Dirichlet and symmetric Logistic Normal distributions for the proportions of total variance $\phi$. Two hyperparameter configurations for $R^2$ were tested: \textit{default} $(\mu_{R^2}, \varphi_{R^2}) = (0.5, 1)$ and \textit{uniform} $(0.5, 2)$. When let $\phi$ follow a symmetric Dirichlet distribution with concentration parameter  $a_\pi \in \{0.5, 1\}$. The \textit{default} case  $a_\pi = 0.5$ encourages shrinkage by pushing mass toward the edges of the simplex, whereas the “\textit{uniform}” case  $a_\pi = 1$ reduces shrinkage. When $a_\pi \leq 0.5$, the marginal prior distributions of the coefficients $b_k$ become unbounded near the origin, enforcing sparsity by shifting mass to the simplex edges \citep{aguilar_intuitive_2023}.

To specify the hyperparameters of $\phi$ when $\phi \sim \LN(\mu, \Sigma)$, we employed KL matching as described in Section \ref{sec:hyperparam_spec}. The outcomes of this specification are referred to as \textit{Full KL}. Specifically: When $a_\pi = 0.5$ we have $\mu_k = 0$, $\sigma_{kk} = \pi$, $\sigma_{kj} = \frac{\sqrt{2}}{2} \pi$.  When $a_\pi = 1$ we have $\mu_k = 0$, $\sigma_{kk} = \frac{\sqrt{3}}{3} \pi$, $\sigma_{kj} = \frac{\sqrt{6}}{6} \pi$. Both cases lead to pairwise correlations of 0.5. Additionally, we explore a variation by restricting $\Sigma^*$ to be a diagonal matrix, referred to as \textit{Scales KL}. Table \ref{tbl:sim_models} summarizes the considered decompositions.

As discussed in Section \ref{sec:hyperparam_spec}, the marginal distributions of $\phi$ for the Logistic Normal (LN) priors may exhibit spikes around $1/K$ as $K$ increases $\mu = 0$. When using diagonal covariance matrices, the LN prior allows for greater variability in the marginals, providing more flexibility and serving as a less informative prior in comparison. If prior knowledge about the location of strong signals exists, this should be encoded in $\mu$. In this study, we mimic complex scenarios with minimal prior knowledge about the coefficients to test the prior under such conditions.

\begin{table}[b]
\begin{center}
\caption{Different decompositions that are considered in the simulations.}
\scalebox{1}{
     \begin{tabular}{lcll}
        \toprule
        Name & Label & Distribution &Details \\
        \midrule
        Dirichlet &  D2  & Dirichlet & $\alpha_k = a_\pi$ \\
        Full KL & LNF & logistic normal & $\mu^*, \Sigma^*$ from KL matching \\
        Scales KL & LNS & logistic normal & $\mu^*, \text{diag}(\Sigma^*)$ from KL matching \\
        \bottomrule
    \end{tabular}
    }
\end{center}
\label{tbl:sim_models}
\end{table}

We crossed the default and uniform hyperparameter specifications for $R^2$ and $a_\pi$, resulting in four hyperparameter specifications. We present a subset of the results for the default-default case $(\mu_{R^2}, \varphi_{R^2}, a_\pi) =(0.5, 1, 0.5)$ and show others in Appendix \ref{appendix}. Additionally, we include results for an informative case in the fixed coefficient setup in Appendix \ref{appendix}, where $\alpha$ is constructed in a way that incorporates strong user knowledge to emphasize important signals, matching the locations of the signals. This helps assess the extent to which both decompositions benefit from user-provided information.

After fully crossing all conditions, we obtained a total of 192 different simulation configurations. For each configuration, we generated $T = 100$ datasets consisting of $N = 100$ training observations $y_n$, $n = 1, \ldots, N$. Predictive metrics (discussed below) were computed based on $N_{\rm test} = 500$ test observations, which were generated independently from the training data using the same data-generating mechanism. The simulations were carried out on the Linux HPC Cluster (LiDO3) at TU Dortmund University.

\subsubsection{Evaluation metrics}
\label{subsec:metrics}

We evaluated and compared the performance of the models based on two criteria: out-of-sample predictive performance and parameter recovery \citep{Vehtari2012Predictive, robert2007bayesian}. Out-of-sample predictive performance was assessed using the expected log-pointwise predictive density ($\elpd$) \citep{Vehtari2012Predictive}, computed on the test data as $\elpd = \sum_{i=1}^N \ln \left( \frac{1}{S} \sum_{s=1}^S p(y_i | \theta^{(s)}) \right)$,
where $\theta^{(s)}$ represents the $s$th posterior draw for $s = 1, \ldots, S$. The $\elpd$ quantifies predictive accuracy across the $N$ data points, evaluating the quality of the overall predictive distribution. A higher $\elpd$ value indicates better out-of-sample predictive performance. Log probability scores, such as $\elpd$, are often recommended as a general-purpose metric when there is no specific reason to use an alternative \citep{Vehtari2012Predictive, VehtariWAIC, vehtariPSIS}.

Parameter recovery was evaluated using the posterior Root Mean-Squared Error (RMSE), calculated as $\rmse = \frac{1}{K} \sum_{k=1}^K \sqrt{\frac{1}{S} \sum_{s=1}^S \left(b_k^{(s)} - b_k \right)^2}$, where $b_k^{(s)}$ represents the $s$th posterior draw for $k = 1, \ldots, K$, and $b_k$ denotes the true value of the $k$th regression coefficient. The RMSE serves as a comprehensive measure of estimation error, naturally capturing the tradeoff between bias and variance. In our analysis, we compute three versions of RMSE: (1) Averaged over all coefficients, (2) only over the truly zero coefficients, (3) only over the truly nonzero coefficients.

To provide a more comprehensive evaluation of the posterior inference properties, we assess the coverage of 95\% marginal credible intervals for each approach based on average interval width, coverage proportion, specificity, and sensitivity. We also present ROC curves. These results are presented in Appendix \ref{appendix}.

To compare multiple models based on a metric of interest $\mathcal{F}$ (e.g., $\elpd$, RMSE), we computed the difference relative to the best-performing model for a given dataset. Let $\{m_1, \ldots, m_l\}$ represent the set of models, and denote the best model with respect to $\mathcal{F}$ by $m^*$. The Delta metric, $\Delta \mathcal{F}_i$, for the $i$th model is defined as $\Delta \mathcal{F}_i = \mathcal{F}(m_i) - \mathcal{F}(m^*)$, where $i = 1, \ldots, l$. This approach highlights differences between models while accounting for variations caused by randomness in the simulated datasets, thus focusing the evaluation on meaningful distinctions.

Metrics for diagnosing and evaluating the MCMC sampler, including Effective Sample Size (ESS) and Rhat \citep{GeyerPracticalMC, handbookmcmc, vehtariRhat}, are provided in Appendix \ref{appendix}. Overall, the results indicate effective sampling from the posterior, as ESS values are sufficiently high and Rhat values fall within acceptable ranges. Additionally, all models demonstrate comparable computational efficiency, with similar runtimes observed across the board.

\subsubsection{Results: Simulated coefficients setup}

We present results for the scenario where the coefficients are generated with an AR(1) correlation structure, characterized by $\rho_b = 0.8$ and marginal variances $\sigma_b^2 = 9$, alongside an average sparsity of $\nu = 0.75$. In real-world applications, it is reasonable to encounter clusters of coefficients that are correlated with each other, even though these correlations do not necessarily translate into significant effects on the response variable.
%----------------------------
\begin{figure}[t]
    \centering
    \includegraphics[width=1\linewidth]{figs/comparison_R2_models/gen_coef_sparse_norm_cor/_delta_lpd_test-gen_coef_sparse_norm_cor.pdf}
    \vspace{-0.5cm}  
    \caption{\textbf{Simulated coefficients setup.} $\Delta \elpd$  evaluated on test datasets of size $N=500$ for the simulated coefficients setup.} 
    \label{fig:delta_lpd_test_sparse_norm}
\end{figure}
%----------------------------
Figure \ref{fig:delta_lpd_test_sparse_norm} shows the distribution of $\Delta \elpd$ across simulations under various conditions, visualized as violin plots with embedded boxplots \citep{violinplots, boxplots}. The results demonstrate that Logistic Normal (LN) decompositions consistently outperform their Dirichlet counterparts. Notably, the LN priors exhibit the best average performance across all scenarios. The LNS version is the clear winner in predictive performance. 
%----------------------------
\begin{figure}[t]
    \centering
    \includegraphics[width=1\linewidth]%{figs/sims_joint_api_0.5/delta_rmse_postbpp_allcases_id_1}
{figs/comparison_R2_models/gen_coef_sparse_norm_cor/delta_rmse_postbpp_allcases_gen_coef_sparse_norm_cor.pdf}
    \vspace{-0.5cm}  
    \caption{\textbf{Simulated coefficients setup.} Violin plots with embedded box plots for $\Delta \rmse$ under different simulation conditions. $\Delta \rmse$ has been partitioned with respect to truly zero and nonzero coefficients. }
    \label{fig:delta_rmse_postbpp_allcases_sparse_norm_cor}
\end{figure}
%----------------------------
Figure \ref{fig:delta_rmse_postbpp_allcases_sparse_norm_cor} depicts the distribution of $\Delta \rmse$ across various simulation conditions. Similar to predictive performance, the LN priors demonstrate the most robust parameter recovery among the competing $R^2$-based models, regardless of the values of $\rho$, $R^2$, or $K$. The overall $\rmse$ results reflect a balance between the model's posterior error for truly zero and nonzero coefficients. To explore this further, Figure \ref{fig:delta_rmse_postbpp_allcases_sparse_norm_cor} separately illustrates $\Delta \rmse$ for truly zero and nonzero coefficients.

These results suggest that LN decompositions achieve more effective shrinkage compared to Dirichlet counterparts, reducing false positives and improving the detection of truly zero coefficients. Dirichlet decompositions impose less shrinkage, which can result in poorer performance when coefficients are truly zero. The trends for nonzero coefficients are similar. Of particular note is the case with $K = 750$, a challenging scenario where shrinkage priors are expected to overshrink. In this high-dimensional setting, LN decompositions show improved behavior when moving from $\rho = 0$ to $\rho = 0.80$. This improvement is also present when transitioning from a low signal ($R^2 = 0.25$) to a moderate signal ($R^2 = 0.60$). Overall, LN decompositions offer a more robust and adaptable approach to shrinkage in sparse, high-dimensional contexts.

In summary, using a zero mean vector and a diagonal covariance matrix for LN priors appears to be an effective strategy, particularly when prior knowledge about relationships between coefficients is limited—a common scenario in practice. These results highlight the potential of LN-based approaches as robust default alternatives to Dirichlet-based shrinkage priors. Even with default parameter settings, LN priors demonstrate competitive or superior performance across diverse scenarios. 

\subsubsection{Results: Fixed coefficients setup}

We show the results when the fixed value is equal to 3. Figure \ref{fig:delta_lpd_test_fixed3} reveals that the LN decompositions consistently excel in out-of-sample predictive performance when compared to the Dirichlet counterpart. The LNS stands out as the most efficient predictive model. The most pronounced disparities
emerge in scenarios characterized by high correlation and dimensionality  ($\rho= 0.8, K=750$), where the performance of the Dirichlet prior is notably subpar.
%----------------------------
\begin{figure}[t]
    \centering
       \includegraphics[width=1\linewidth]{figs/comparison_R2_models/gen_coef_fixed3/_delta_lpd_test-gen_coef_fixed3.pdf}
    \vspace{-0.5cm}  
    \caption{\textbf{Fixed coefficients setup.} $\Delta \elpd$ evaluated on the test datasets for the fixed coefficients when the signal have a value of 3.}
    \label{fig:delta_lpd_test_fixed3}
\end{figure}
%---------------------------
The results for $\Delta \rmse$ for all of the coefficients and the zero coefficients behave similarly to the simulated coefficients setup. The LN decompositions perform clearly and uniformly better the Dirichlet for all the coefficients and for the zero coefficient only. This indicates again that LN decompositions are better at both overall $\rmse$ and noise detection. We only show the case for nonzero coefficients in Figure \ref{fig:delta_rmse_postbppn0_fixed3}. A complete overview over the other cases is available in Appendix \ref{appendix}. 

The Dirichlet decomposition excels only in the in low-signal, high-correlation settings. LN decompositions achieves better performance in all other scenarios. These results suggest that the effectiveness of  LN decompositions could be enhanced even more by incorporating prior structure that mitigates correlations in the design matrix. If we focus on the $\rho= 0.8, K=750$ case in both Figures \ref{fig:delta_lpd_test_fixed3} and \ref{fig:delta_rmse_postbppn0_fixed3}, we can see that there is a tradeoff between out-of-sample predictive performance and detecting signals in that case. Here, the Dirichlet is not shrinking sufficiently (leading to worse out-of-sample predictions), while the LN might be overshrinking some signals (but ensures better out-of-sample predictions overall).
%----------------------------
\begin{figure}[bt]
\includegraphics[width=1\linewidth]{figs/comparison_R2_models/gen_coef_fixed3/_delta_rmse_bppn0-gen_coef_fixed3.pdf}
    \caption{\textbf{Fixed coefficients setup.} $\Delta \rmse$ for truly nonzero coefficients in the fixed coefficients case setup.}
    \label{fig:delta_rmse_postbppn0_fixed3}
\end{figure}
%----------------------------

\subsection{Real-world case studies}

\begin{figure}[t!]
    \centering
    \includegraphics[width=0.95\linewidth]{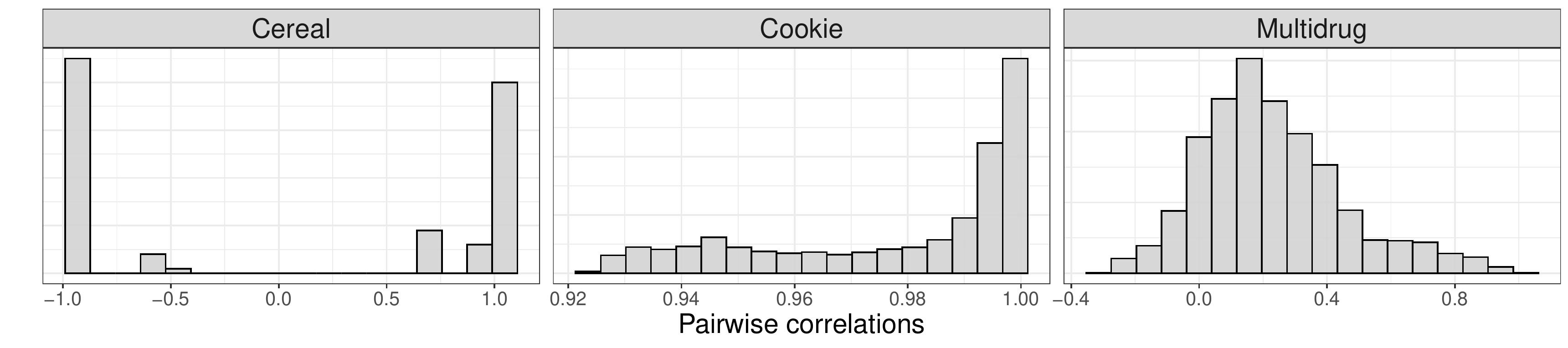}
    \caption{Histograms of pairwise correlations of the predictor variables of the different datasets considered in the case studies.}
    \label{fig:pairwisecors}
\end{figure}

We evaluate the predictive performance of our proposed prior using three high dimensional real-world datasets, each with distinct correlation structures among covariates. The datasets are as follows: 1) \textit{Cereal:} Comprises starch content measurements from 15 observations with 145 infrared spectra measurements as predictors, sourced from the R package \texttt{chemometrics} \citep{chemometrics}. 2)  \textit{Cookie:} Originating from a study on near-infrared (NIR) spectroscopy of biscuit dough, this dataset includes fat content measurements across 72 samples with 700 NIR spectra predictors. It was first introduced by \cite{CookiesOsborne1984ApplicationON} and is available in the orphaned \texttt{ppls} R package \citep{ppls}. 3) \textit{Multidrug:} Arises from a pharmacogenomic study investigating the relationship between drug concentration (at which 50\% growth is inhibited for a human cell line) and the expression of the adenosine triphosphate-binding cassette transporter \citep{SZAKACS2004129}. After removing missing values, it includes 853 covariates and 60 observations. The Cereal and Multidrug datasets have previously been studied in the context of shrinkage priors by \cite{PolsonLevy, Griffinsomepriors, r2d2zhang}, while the Cookie dataset has been analyzed by \cite{CookiesOsborne1984ApplicationON, Ghoshvariableselection, r2d2zhang}. These datasets exhibit distinct dependency structures among predictor variables: a mix of positive and negative correlations (Cereal) to strongly positively correlated (Cookie) and low to moderate pairwise correlations (Multidrug) \citep{r2d2zhang}. We represent this via the histograms of pairwise correlations in Figure \ref{fig:pairwisecors}.

%----------------------------
%----------------------------
We use $\elpd$ to quantify out-of-sample predictive performance. Since we don't have an independent test set available, we use cross-validation to estimate out-of-sample ELPD as follows: We split each dataset into training and test sets, with 75\% of observations used for training and the remaining 25\% for testing. We repeat this 100 times and use the resulting $\elpd$ values, which are subsequently summed to obtain an overall $\elpd$ estimate. We compare the performance of the Dirichlet, LNF, and LNS decompositions with specific parameter settings, namely $(\mu_{R^2}, \varphi_{R^2}, a_\pi)= (1, 0.5, 0.5)$. 

\begin{table}[H]
\caption{Differences in $\elpd$ and standard deviations for the different datasets considered, computed through pairwise comparisons with the model having the highest $\elpd$ (in the first row). The initial value is zero, and subsequent rows display negative values, indicating the difference with the best model \citep{looR}. See \cite{VehtariWAIC} for details on standard error calculations.}
\label{tab:data_1}
\begin{threeparttable}
\resizebox{\textwidth}{!}{%
\begin{tabular}{@{}lccc@{}}
\toprule
Dataset     & Cereal ($N=15, K=145$) & Cookie ($N=72, K=700$) & Multidrug ($N=60, K=853$) \\ 
\midrule
Model       & $\Delta \elpd$ (SE)      & $\Delta \elpd$ (SE)      & $\Delta \elpd$ (SE)         \\ 
\midrule
LNS         & 0.0 (0.0)             & 0.0 (0.0)             & 0.0 (0.0)                \\
LNF         & -101.4 (85.1)         & -673.1 (471.5)        & -2950.4 (743.8)          \\
D2         & -96.8 (42.2)          & -518.9 (274.4)        & -5464.2 (794.7)          \\
\bottomrule
\end{tabular}%
}
\end{threeparttable}
\end{table}

%$\elpd$ results indicate a better performance for the LN decompositions, specifically of LNS, over Dirichlet counterparts, aligning with simulation results.
We summarise the results in Table \ref{tab:data_1}. LNS consistently demonstrates superior $\elpd$ performance compared to both Dirichlet and LNF across the diverse datasets. The latter two priors clearly differ from each other only for the Multidrug data where LNF was better than Dirichlet. Together, this showcases that priors that exert more shrinkage on complex, high-dimensional datasets can be beneficial in terms of predictive performance.

\section{Discussion}

We introduced the Generalized Decomposition R2 prior framework for high-dimensional Bayesian regression, building on and extending the R2D2 prior \citep{r2d2zhang, aguilar_intuitive_2023}. Our key innovation lies in allowing to vary the prior used for the explained variance decomposition, enabling a more nuanced exploration of dependency structures. Previous methods are confined to Dirichlet decompositions, and thus can only express negative covariances tied to the means of the explained variance proportions. We overcome this limitation by employing the logistic normal distribution, facilitating the expression of covariance structures by transitioning from the simplex to the unconstrained real space.

Our simulations and real-world case studies demonstrate substantial gains in predictive performance when using logistic normal priors for the proportions of explained variance. The flexible dependency structures encoded in the logistic normal allow for a broader range of prior assumptions compared to the Dirichlet distribution. Specifically, we show that use of logistic normal based decompositions leads to superior out-of-sample predictive performance compared to the use of Dirichlet counterparts. In terms of parameter recovery, the Logistic normal prior models showed on par and superior performance compared to Dirichlet in most scenarios.

Our primary goal was to build well-predicting models, using the full set of available covariates and strong shrinkage on the corresponding coefficients. That said, as is the case for continuous shrinkage priors more generally, they do not create exact sparsity directly. Rather, redundant coefficients are shrunk to values close to zero, but not to zero exactly. As a result, predictive capabilities are not necessarily directly visible in few nonzero coefficients but rather implicitly distributed also among the almost zero coefficients, at least in high-dimensional settings \citep{PiironenHorseshoe}. In other words, we argue that our models (and other continuous shrinkage priors models) should not be used directly for variable selection (for more details and discussion see \cite{PiironenProjInf, ZhangCredibleRegions, bayesian_variable_selection_handbook}). Rather, we recommend a two-step procedure where, after fitting a well predicting reference model with all covariates in the first step, we apply a separate, dedicated variable selection procedure in the second step, for example, projection predictive variable selection \citep{PiironenProjInf, catalina2020projection, projpredmclatchie2023robust}.

In future research, one important direction will be to study the influence and optimal choice of the logistic normal hyperparameters in more detail. While the hyperparameter choices proposed here already showed strongly improved performance compared to Dirichlet, it is still unclear how much room for improvement remains within the class of logistic normal decomposition priors. One promising direction we see in this context is the inclusion of covariate grouping or covariate dependency information in the prior \citep{BossGIGG}. But exactly how this information should be encoded, and if its inclusion is actually beneficial for the prior's performance, remains to be studied.

% ** Acknowledgements **
\begin{acks}[Acknowledgments]
Funded by Deutsche Forschungsgemeinschaft (DFG, German Research Foundation) under Germany's Excellence Strategy EXC 2075-390740016 and DFG Project 500663361. We acknowledge the support by the Stuttgart Center for Simulation Science (SimTech). We acknowledge the computing time
provided on the Linux HPC cluster at Technical University
Dortmund (LiDO3), partially funded in the course of the
Large-Scale Equipment Initiative by DFG Project 271512359. 
\end{acks}

% ** The bibliograhy **\

\bibliographystyle{ba}
\bibliography{bib/refs}

\begin{thebibliography}{103}
\newcommand{\enquote}[1]{``#1''}
\expandafter\ifx\csname natexlab\endcsname\relax\def\natexlab#1{#1}\fi
\expandafter\ifx\csname url\endcsname\relax
  \def\url#1{{\tt #1}}\fi
\expandafter\ifx\csname urlprefix\endcsname\relax\def\urlprefix{URL }\fi
\ifx\endbibitem\undefined \let\endbibitem\relax\fi

\bibitem[{Agresti and Hitchcock(2005)}]{CATDAagresti_bayesian_2005}
Agresti, A. and Hitchcock, D.~B. (2005).
\newblock \enquote{Bayesian inference for categorical data analysis.}
\newblock {\em Statistical Methods and Applications\/}, 14(3): 297--330.
\newline\urlprefix\url{https://doi.org/10.1007/s10260-005-0121-y}
\endbibitem

\bibitem[{Aguilar and B{\"u}rkner(2023)}]{aguilar_intuitive_2023}
Aguilar, J.~E. and B{\"u}rkner, P.-C. (2023).
\newblock \enquote{{Intuitive joint priors for Bayesian linear multilevel
  models: The R2D2M2 prior}.}
\newblock {\em Electronic Journal of Statistics\/}, 17(1): 1711 -- 1767.
\newline\urlprefix\url{https://doi.org/10.1214/23-EJS2136}
\endbibitem

\bibitem[{Aguilar and Bürkner(2025)}]{aguilar2024suppl}
Aguilar, J.~E. and Bürkner, P.-C. (2025).
\newblock \enquote{Supplement to "Generalized Decomposition Priors on R2".}
\newblock {\em Bayesian Analysis\/}.
\endbibitem

\bibitem[{Aitchison and Shen(1980)}]{logisticnormal}
Aitchison, J. and Shen, S.~M. (1980).
\newblock \enquote{Logistic-Normal Distributions: Some Properties and Uses.}
\newblock {\em Biometrika\/}, 67(2): 261--272.
\newline\urlprefix\url{http://www.jstor.org/stable/2335470}
\endbibitem

\bibitem[{Aitchison(1986)}]{AitchisonMonograph}
Aitchison, J.~J. (1986).
\newblock {\em The statistical analysis of compositional data / J.
  Aitchison.\/}.
\newblock Monographs on statistics and applied probability (Series). Chapman
  and Hall.
\endbibitem

\bibitem[{Armagan et~al.(2011)Armagan, Clyde, and
  Dunson}]{armagan2011generalized}
Armagan, A., Clyde, M., and Dunson, D.~B. (2011).
\newblock \enquote{{Generalized beta mixtures of Gaussians}.}
\newblock In {\em {Advances in neural information processing systems}\/},
  523--531.
\endbibitem

\bibitem[{Armagan et~al.(2013)Armagan, Dunson, and
  Lee}]{armagan2013generalized}
Armagan, A., Dunson, D.~B., and Lee, J. (2013).
\newblock \enquote{{Generalized double Pareto shrinkage}.}
\newblock {\em Statistica Sinica\/}, 23(1): 119.
\endbibitem

\bibitem[{Bai and Ghosh(2019)}]{BaiHypothesisNB}
Bai, R. and Ghosh, M. (2019).
\newblock \enquote{Large-scale multiple hypothesis testing with the normal-beta
  prime prior.}
\newblock {\em Statistics\/}, 53(6): 1210--1233.
\newline\urlprefix\url{https://doi.org/10.1080/02331888.2019.1662017}
\endbibitem

\bibitem[{Barndorff-Nielsen and Jørgensen(1991)}]{simplex_nielsen}
Barndorff-Nielsen, O. and Jørgensen, B. (1991).
\newblock \enquote{Some parametric models on the simplex.}
\newblock {\em Journal of Multivariate Analysis\/}, 39(1): 106--116.
\newline\urlprefix\url{https://www.sciencedirect.com/science/article/pii/0047259X9190008P}
\endbibitem

\bibitem[{Bhadra et~al.(2017)Bhadra, Datta, Polson, and
  Willard}]{Horseshoeplus}
Bhadra, A., Datta, J., Polson, N.~G., and Willard, B. (2017).
\newblock \enquote{{The Horseshoe+ Estimator of Ultra-Sparse Signals}.}
\newblock {\em Bayesian Analysis\/}, 12(4): 1105 -- 1131.
\newline\urlprefix\url{https://doi.org/10.1214/16-BA1028}
\endbibitem

\bibitem[{Bhadra et~al.(2019)Bhadra, Datta, Polson, and
  Willard}]{PolsonHSmeetsLasso}
--- (2019).
\newblock \enquote{{Lasso Meets Horseshoe: A Survey}.}
\newblock {\em Statistical Science\/}, 34(3): 405 -- 427.
\newline\urlprefix\url{https://doi.org/10.1214/19-STS700}
\endbibitem

\bibitem[{Bhattacharya et~al.(2015)Bhattacharya, Pati, Pillai, and
  Dunson}]{DirichletLaplace}
Bhattacharya, A., Pati, D., Pillai, N.~S., and Dunson, D.~B. (2015).
\newblock \enquote{Dirichlet–Laplace Priors for Optimal Shrinkage.}
\newblock {\em Journal of the American Statistical Association\/}, 110(512):
  1479--1490.
\newblock PMID: 27019543.
\newline\urlprefix\url{https://doi.org/10.1080/01621459.2014.960967}
\endbibitem

\bibitem[{Bishop(2006)}]{bishop2006pattern}
Bishop, C. (2006).
\newblock {\em Pattern Recognition and Machine Learning\/}.
\newblock Information Science and Statistics. Springer.
\newline\urlprefix\url{https://books.google.de/books?id=kTNoQgAACAAJ}
\endbibitem

\bibitem[{Blei et~al.(2017)Blei, Kucukelbir, and McAuliffe}]{BleiVI}
Blei, D.~M., Kucukelbir, A., and McAuliffe, J.~D. (2017).
\newblock \enquote{Variational Inference: A Review for Statisticians.}
\newblock {\em Journal of the American Statistical Association\/}, 112(518):
  859–877.
\newline\urlprefix\url{http://dx.doi.org/10.1080/01621459.2017.1285773}
\endbibitem

\bibitem[{Blei and Lafferty(2007)}]{CorTM}
Blei, D.~M. and Lafferty, J.~D. (2007).
\newblock \enquote{{A correlated topic model of Science}.}
\newblock {\em The Annals of Applied Statistics\/}, 1(1): 17 -- 35.
\newline\urlprefix\url{https://doi.org/10.1214/07-AOAS114}
\endbibitem

\bibitem[{Boogaart and Tolosana-Delgado(2013)}]{CompositionsR}
Boogaart, K. and Tolosana-Delgado, R. (2013).
\newblock {\em Analyzing Compositional Data with R\/}, 209--253.
\endbibitem

\bibitem[{Boulesteix(2014)}]{ppls}
Boulesteix, N. K. A.-L. (2014).
\newblock {\em ppls: Penalized Partial Least Squares\/}.
\newblock R package version 1.6-1.
\newline\urlprefix\url{http://cran.nexr.com/web/packages/ppls/index.html}
\endbibitem

\bibitem[{Brooks et~al.(2011)Brooks, Gelman, and Jones}]{handbookmcmc}
Brooks, Gelman, S., and Jones, A. (2011).
\newblock {\em Handbook of Markov Chain Monte Carlo\/}.
\newblock Chapman and Hall/CRC, 1 edition.
\endbibitem

\bibitem[{Bürkner(2017)}]{brmsJSS}
Bürkner, P.-C. (2017).
\newblock \enquote{brms: An R Package for Bayesian Multilevel Models Using
  Stan.}
\newblock {\em Journal of Statistical Software\/}, 80(1): 1–28.
\newline\urlprefix\url{https://www.jstatsoft.org/index.php/jss/article/view/v080i01}
\endbibitem

\bibitem[{Carpenter et~al.(2017)Carpenter, Gelman, Hoffman, Lee, Goodrich,
  Betancourt, Brubaker, Guo, Li, and Riddell}]{StanJSS}
Carpenter, B., Gelman, A., Hoffman, M.~D., Lee, D., Goodrich, B., Betancourt,
  M., Brubaker, M., Guo, J., Li, P., and Riddell, A. (2017).
\newblock \enquote{Stan: A Probabilistic Programming Language.}
\newblock {\em Journal of Statistical Software\/}, 76(1): 1–32.
\newline\urlprefix\url{https://www.jstatsoft.org/index.php/jss/article/view/v076i01}
\endbibitem

\bibitem[{Carvalho et~al.(2010)Carvalho, Polson, and Scott}]{Horseshoe}
Carvalho, Polson, and Scott (2010).
\newblock \enquote{The horseshoe estimator for sparse signals.}
\newblock {\em Biometrika\/}, 97(2): 465--480.
\newline\urlprefix\url{http://www.jstor.org/stable/25734098}
\endbibitem

\bibitem[{Castillo and van~der Vaart(2012)}]{Castillo}
Castillo, I. and van~der Vaart, A. (2012).
\newblock \enquote{{Needles and Straw in a Haystack: Posterior concentration
  for possibly sparse sequences}.}
\newblock {\em The Annals of Statistics\/}, 40(4): 2069 -- 2101.
\newline\urlprefix\url{https://doi.org/10.1214/12-AOS1029}
\endbibitem

\bibitem[{Chen et~al.(2013)Chen, Zhu, Wang, Zheng, and Zhang}]{ScalableCTMLN}
Chen, J., Zhu, J., Wang, Z., Zheng, X., and Zhang, B. (2013).
\newblock \enquote{Scalable Inference for Logistic-Normal Topic Models.}
\newblock In Burges, C., Bottou, L., Welling, M., Ghahramani, Z., and
  Weinberger, K. (eds.), {\em Advances in Neural Information Processing
  Systems\/}, volume~26. Curran Associates, Inc.
\newline\urlprefix\url{https://proceedings.neurips.cc/paper_files/paper/2013/file/285f89b802bcb2651801455c86d78f2a-Paper.pdf}
\endbibitem

\bibitem[{Chow(2022)}]{chow2022schlomilchintegralsprobabilitydistributions}
Chow, D. D.~K. (2022).
\newblock \enquote{Schl\"omilch integrals and probability distributions on the
  simplex.}
\newline\urlprefix\url{https://arxiv.org/abs/2201.11013}
\endbibitem

\bibitem[{Connor and Mosimann(1969)}]{ConnorMossimanGenDirichlet}
Connor, R.~J. and Mosimann, J.~E. (1969).
\newblock \enquote{Concepts of Independence for Proportions with a
  Generalization of the Dirichlet Distribution.}
\newblock {\em Journal of the American Statistical Association\/}, 64(325):
  194--206.
\newline\urlprefix\url{http://www.jstor.org/stable/2283728}
\endbibitem

\bibitem[{Creus-Mart{\'i} et~al.(2021)Creus-Mart{\'i}, Moya, and
  Santonja}]{CreusMart2021ADA}
Creus-Mart{\'i}, I., Moya, A., and Santonja, F.-J. (2021).
\newblock \enquote{A Dirichlet Autoregressive Model for the Analysis of
  Microbiota Time-Series Data.}
\newblock {\em Complex.\/}, 2021: 9951817:1--9951817:16.
\newline\urlprefix\url{https://api.semanticscholar.org/CorpusID:237715479}
\endbibitem

\bibitem[{Efron(2011)}]{TweedieEfron}
Efron, B. (2011).
\newblock \enquote{Tweedie’s Formula and Selection Bias.}
\newblock {\em Journal of the American Statistical Association\/}, 106(496):
  1602--1614.
\newblock PMID: 22505788.
\newline\urlprefix\url{https://doi.org/10.1198/jasa.2011.tm11181}
\endbibitem

\bibitem[{Filzmoser(2023)}]{chemometrics}
Filzmoser, P. (2023).
\newblock {\em chemometrics: Multivariate Statistical Analysis in
  Chemometrics\/}.
\newblock R package version 1.4.4.
\newline\urlprefix\url{https://CRAN.R-project.org/package=chemometrics}
\endbibitem

\bibitem[{Follett and Yu(2019)}]{FOLLETT2019130}
Follett, L. and Yu, C. (2019).
\newblock \enquote{Achieving parsimony in Bayesian vector autoregressions with
  the horseshoe prior.}
\newblock {\em Econometrics and Statistics\/}, 11: 130--144.
\newline\urlprefix\url{https://www.sciencedirect.com/science/article/pii/S2452306219300036}
\endbibitem

\bibitem[{Frederic and Lad(2008)}]{MomentsLogitNormal_2}
Frederic, P. and Lad, F. (2008).
\newblock \enquote{Two Moments of the Logitnormal Distribution.}
\newblock {\em Communications in Statistics - Simulation and Computation\/},
  37(7): 1263--1269.
\newline\urlprefix\url{https://doi.org/10.1080/03610910801983178}
\endbibitem

\bibitem[{Gelman(2006{\natexlab{a}})}]{GelmanCauchy}
Gelman, A. (2006{\natexlab{a}}).
\newblock \enquote{Prior Distributions for Variance Parameters in Hierarchical
  Models.}
\newblock {\em Bayesian Analysis\/}, 1.
\endbibitem

\bibitem[{Gelman(2006{\natexlab{b}})}]{GelmanHalfStudentt}
--- (2006{\natexlab{b}}).
\newblock \enquote{{Prior distributions for variance parameters in hierarchical
  models (comment on article by Browne and Draper)}.}
\newblock {\em Bayesian Analysis\/}, 1(3): 515 -- 534.
\newline\urlprefix\url{https://doi.org/10.1214/06-BA117A}
\endbibitem

\bibitem[{Gelman et~al.(2013)Gelman, Carlin, Stern, Dunson, Vehtari, and
  Rubin}]{gelman2013bda}
Gelman, A., Carlin, J., Stern, H., Dunson, D., Vehtari, A., and Rubin, D.
  (2013).
\newblock {\em Bayesian Data Analysis, Third Edition\/}.
\newblock Chapman \& Hall/CRC Texts in Statistical Science. Taylor \& Francis.
\newline\urlprefix\url{https://books.google.de/books?id=ZXL6AQAAQBAJ}
\endbibitem

\bibitem[{Gelman and Hill(2006)}]{gelman_hill_2006}
Gelman, A. and Hill, J. (2006).
\newblock {\em Data Analysis Using Regression and Multilevel/Hierarchical
  Models\/}.
\newblock Analytical Methods for Social Research. Cambridge University Press.
\endbibitem

\bibitem[{Gelman et~al.(2020)Gelman, Hill, and Vehtari}]{gelmanregstories2020}
Gelman, A., Hill, J., and Vehtari, A. (2020).
\newblock {\em Regression and Other Stories\/}.
\newblock Analytical Methods for Social Research. Cambridge University Press.
\endbibitem

\bibitem[{Geyer(1992)}]{GeyerPracticalMC}
Geyer, C.~J. (1992).
\newblock \enquote{{Practical Markov Chain Monte Carlo}.}
\newblock {\em Statistical Science\/}, 7(4): 473 -- 483.
\newline\urlprefix\url{https://doi.org/10.1214/ss/1177011137}
\endbibitem

\bibitem[{Ghosal et~al.(2000)Ghosal, Ghosh, and van~der
  Vaart}]{ghosalconvergence}
Ghosal, S., Ghosh, J.~K., and van~der Vaart, A.~W. (2000).
\newblock \enquote{{Convergence rates of posterior distributions}.}
\newblock {\em The Annals of Statistics\/}, 28(2): 500 -- 531.
\newline\urlprefix\url{https://doi.org/10.1214/aos/1016218228}
\endbibitem

\bibitem[{Ghosh and Ghattas(2015)}]{Ghoshvariableselection}
Ghosh, J. and Ghattas, A. (2015).
\newblock \enquote{Bayesian Variable Selection Under Collinearity.}
\newblock {\em The American Statistician\/}, 69.
\endbibitem

\bibitem[{Ghosh et~al.(2019)Ghosh, Yao, and Doshi-Velez}]{Ghosh2019}
Ghosh, S., Yao, J., and Doshi-Velez, F. (2019).
\newblock \enquote{Model Selection in Bayesian Neural Networks via Horseshoe
  Priors.}
\newblock {\em Journal of Machine Learning Research\/}, 20(182): 1--46.
\newline\urlprefix\url{http://jmlr.org/papers/v20/19-236.html}
\endbibitem

\bibitem[{Giraud(2014)}]{giraud2014HD}
Giraud, C. (2014).
\newblock {\em Introduction to High-Dimensional Statistics\/}.
\newblock Chapman \& Hall/CRC Monographs on Statistics \& Applied Probability.
  Taylor \& Francis.
\newline\urlprefix\url{https://books.google.de/books?id=qRuVoAEACAAJ}
\endbibitem

\bibitem[{Good(1962)}]{Jeffreys}
Good, I.~J. (1962).
\newblock \enquote{Theory of Probability Harold Jeffreys (Third edition, 447 +
  ix pp., Oxford Univ. Press, 84s.).}
\newblock {\em Geophysical Journal International\/}, 6: 555--558.
\endbibitem

\bibitem[{Goodfellow et~al.(2016)Goodfellow, Bengio, and
  Courville}]{Goodfellow}
Goodfellow, I.~J., Bengio, Y., and Courville, A. (2016).
\newblock {\em Deep Learning\/}.
\newblock Cambridge, MA, USA: MIT Press.
\newblock \url{http://www.deeplearningbook.org}.
\endbibitem

\bibitem[{Goodrich et~al.(2020)Goodrich, Gabry, Ali, and Brilleman}]{rstanarm}
Goodrich, B., Gabry, J., Ali, I., and Brilleman, S. (2020).
\newblock \enquote{rstanarm: {Bayesian} applied regression modeling via
  {Stan}.}
\newblock R package version 2.21.1.
\newline\urlprefix\url{https://mc-stan.org/rstanarm}
\endbibitem

\bibitem[{Greenacre et~al.(2023)Greenacre, Grunsky, Bacon-Shone, Erb, and
  Quinn}]{Aitchison40years}
Greenacre, M., Grunsky, E., Bacon-Shone, J., Erb, I., and Quinn, T. (2023).
\newblock \enquote{{Aitchison’s Compositional Data Analysis 40 Years on: A
  Reappraisal}.}
\newblock {\em Statistical Science\/}, 38(3): 386 -- 410.
\newline\urlprefix\url{https://doi.org/10.1214/22-STS880}
\endbibitem

\bibitem[{Griffin and Brown(2010)}]{griffin2010inference}
Griffin, J.~E. and Brown, P.~J. (2010).
\newblock \enquote{{Inference with normal-gamma prior distributions in
  regression problems}.}
\newblock {\em Bayesian Analysis\/}, 5(1): 171--188.
\endbibitem

\bibitem[{Griffin and Brown(2013)}]{Griffinsomepriors}
--- (2013).
\newblock \enquote{{Some Priors for Sparse Regression Modelling}.}
\newblock {\em Bayesian Analysis\/}, 8(3): 691 -- 702.
\newline\urlprefix\url{https://doi.org/10.1214/13-BA827}
\endbibitem

\bibitem[{Gupta and Richards(2001)}]{Dirichlet_history}
Gupta, R.~D. and Richards, D. S.~P. (2001).
\newblock \enquote{The History of the Dirichlet and Liouville Distributions.}
\newblock {\em International Statistical Review / Revue Internationale de
  Statistique\/}, 69(3): 433--446.
\newline\urlprefix\url{http://www.jstor.org/stable/1403455}
\endbibitem

\bibitem[{Hastie et~al.(2015)Hastie, Tibshirani, and
  Wainwright}]{hastie2015statisticalsparsity}
Hastie, T., Tibshirani, R., and Wainwright, M. (2015).
\newblock {\em Statistical Learning with Sparsity: The Lasso and
  Generalizations\/}.
\newblock ISSN. CRC Press.
\newline\urlprefix\url{https://books.google.de/books?id=f-A_CQAAQBAJ}
\endbibitem

\bibitem[{Hintze and Nelson(1998)}]{violinplots}
Hintze, J.~L. and Nelson, R.~D. (1998).
\newblock \enquote{Violin Plots: A Box Plot-Density Trace Synergism.}
\newblock {\em The American Statistician\/}, 52(2): 181--184.
\newline\urlprefix\url{https://www.tandfonline.com/doi/abs/10.1080/00031305.1998.10480559}
\endbibitem

\bibitem[{Hoerl and Kennard(1970)}]{Ridge}
Hoerl, A.~E. and Kennard, R.~W. (1970).
\newblock \enquote{Ridge Regression: Biased Estimation for Nonorthogonal
  Problems.}
\newblock {\em Technometrics\/}, 12(1): 55--67.
\newline\urlprefix\url{http://www.jstor.org/stable/1267351}
\endbibitem

\bibitem[{Hoffman and Gelman(2011)}]{hoffman2011nouturn}
Hoffman, M.~D. and Gelman, A. (2011).
\newblock \enquote{The No-U-Turn Sampler: Adaptively Setting Path Lengths in
  Hamiltonian Monte Carlo.}
\endbibitem

\bibitem[{Holmes and Schofield(2022)}]{Momentslogitnormal}
Holmes, J.~B. and Schofield, M.~R. (2022).
\newblock \enquote{Moments of the logit-normal distribution.}
\newblock {\em Communications in Statistics - Theory and Methods\/}, 51(3):
  610--623.
\newline\urlprefix\url{https://doi.org/10.1080/03610926.2020.1752723}
\endbibitem

\bibitem[{Jeffrey et~al.(2007)Jeffrey, Zwillinger, Gradshteyn, and
  Ryzhik}]{Zwillinger}
Jeffrey, A., Zwillinger, D., Gradshteyn, I., and Ryzhik, I. (2007).
\newblock \enquote{8–9 - Special Functions.}
\newblock In {\em Table of Integrals, Series, and Products (Seventh
  Edition)\/}, 859--1048. Boston: Academic Press, seventh edition edition.
\newline\urlprefix\url{https://www.sciencedirect.com/science/article/pii/B9780080471112500169}
\endbibitem

\bibitem[{Johndrow et~al.(2020)Johndrow, Orenstein, and
  Bhattacharya}]{ScalableHSBhattacharya}
Johndrow, J., Orenstein, P., and Bhattacharya, A. (2020).
\newblock \enquote{Scalable Approximate MCMC Algorithms for the Horseshoe
  Prior.}
\newblock {\em Journal of Machine Learning Research\/}, 21(73): 1--61.
\newline\urlprefix\url{http://jmlr.org/papers/v21/19-536.html}
\endbibitem

\bibitem[{Kohns and Szendrei(2024)}]{Kohns2020HorseshoePB}
Kohns, D. and Szendrei, T. (2024).
\newblock \enquote{Horseshoe prior Bayesian quantile regression.}
\newblock {\em Journal of the Royal Statistical Society Series C: Applied
  Statistics\/}, 73(1): 193--220.
\endbibitem

\bibitem[{Kruijer et~al.(2010)Kruijer, Rousseau, and Vaart}]{SimplexKruijer}
Kruijer, W., Rousseau, J., and Vaart, A. (2010).
\newblock \enquote{Adaptive Bayesian Density Estimation with Location-Scale
  Mixtures.}
\newblock {\em Electronic Journal of Statistics\/}, 4.
\endbibitem

\bibitem[{Kruschke(2015)}]{meanprecbeta}
Kruschke, J.~K. (2015).
\newblock \enquote{Chapter 6 - Inferring a Binomial Probability via Exact
  Mathematical Analysis.}
\newblock In Kruschke, J.~K. (ed.), {\em Doing Bayesian Data Analysis (Second
  Edition)\/}, 123--141. Boston: Academic Press, second edition edition.
\newline\urlprefix\url{https://www.sciencedirect.com/science/article/pii/B9780124058880000064}
\endbibitem

\bibitem[{Kullback and Leibler(1951)}]{KLdiv}
Kullback, S. and Leibler, R.~A. (1951).
\newblock \enquote{{On Information and Sufficiency}.}
\newblock {\em The Annals of Mathematical Statistics\/}, 22(1): 79 -- 86.
\newline\urlprefix\url{https://doi.org/10.1214/aoms/1177729694}
\endbibitem

\bibitem[{Leydold and Hörmann(2011)}]{GenInvGaussian}
Leydold, J. and Hörmann, W. (2011).
\newblock \enquote{Generating generalized inverse Gaussian random variates by
  fast inversion.}
\newblock {\em Computational Statistics and Data Analysis\/}, 55(1): 213--217.
\newline\urlprefix\url{https://www.sciencedirect.com/science/article/pii/S0167947310002847}
\endbibitem

\bibitem[{Lin(2016)}]{OnTheDirichlet}
Lin, J. (2016).
\newblock \enquote{On The Dirichlet Distribution by Jiayu Lin.}
\endbibitem

\bibitem[{Mikkola et~al.(2021)Mikkola, Martin, Chandramouli, Hartmann, Pla,
  Thomas, Pesonen, Corander, Vehtari, Kaski, Bürkner, and Klami}]{prioreli}
Mikkola, P., Martin, O.~A., Chandramouli, S., Hartmann, M., Pla, O.~A., Thomas,
  O., Pesonen, H., Corander, J., Vehtari, A., Kaski, S., Bürkner, P.-C., and
  Klami, A. (2021).
\newblock \enquote{Prior knowledge elicitation: The past, present, and future.}
\endbibitem

\bibitem[{Mimno et~al.(2008)Mimno, Wallach, and McCallum}]{CTMmimno2008gibbs}
Mimno, D., Wallach, H., and McCallum, A. (2008).
\newblock \enquote{Gibbs sampling for logistic normal topic models with
  graph-based priors.}
\endbibitem

\bibitem[{Neal(2003)}]{neal_slice_2003}
Neal, R.~M. (2003).
\newblock \enquote{Slice sampling.}
\newblock {\em The Annals of Statistics\/}, 31(3): 705--767.
\newblock Publisher: Institute of Mathematical Statistics.
\newline\urlprefix\url{https://projecteuclid.org/journals/annals-of-statistics/volume-31/issue-3/Slice-sampling/10.1214/aos/1056562461.full}
\endbibitem

\bibitem[{Ongaro and Migliorati(2013)}]{simplex_ongaro}
Ongaro, A. and Migliorati, S. (2013).
\newblock \enquote{A generalization of the Dirichlet distribution.}
\newblock {\em Journal of Multivariate Analysis\/}, 114: 412--426.
\newline\urlprefix\url{https://www.sciencedirect.com/science/article/pii/S0047259X12001753}
\endbibitem

\bibitem[{Osborne et~al.(1984)Osborne, Fearn, Miller, and
  Douglas}]{CookiesOsborne1984ApplicationON}
Osborne, B., Fearn, T., Miller, A.~R., and Douglas, S. (1984).
\newblock \enquote{Application of near infrared reflectance spectroscopy to the
  compositional analysis of biscuits and biscuit doughs.}
\newblock {\em Journal of the Science of Food and Agriculture\/}, 35: 99--105.
\newline\urlprefix\url{https://api.semanticscholar.org/CorpusID:94286061}
\endbibitem

\bibitem[{Pavone et~al.(2020)Pavone, Piironen, Bürkner, and
  Vehtari}]{pavone2020using}
Pavone, F., Piironen, J., Bürkner, P.-C., and Vehtari, A. (2020).
\newblock \enquote{Using reference models in variable selection.}
\endbibitem

\bibitem[{Pawlowsky-Glahn and Buccianti(2011)}]{PawlowskyCompositionalDA}
Pawlowsky-Glahn, V. and Buccianti, A. (2011).
\newblock \enquote{Compositional data analysis : theory and applications.}
\newline\urlprefix\url{https://api.semanticscholar.org/CorpusID:117956959}
\endbibitem

\bibitem[{Piironen et~al.(2020)Piironen, Paasiniemi, and
  Vehtari}]{PiironenProjInf}
Piironen, J., Paasiniemi, M., and Vehtari, A. (2020).
\newblock \enquote{{Projective inference in high-dimensional problems:
  Prediction and feature selection}.}
\newblock {\em Electronic Journal of Statistics\/}, 14(1): 2155 -- 2197.
\newline\urlprefix\url{https://doi.org/10.1214/20-EJS1711}
\endbibitem

\bibitem[{Piironen and Vehtari(2017)}]{PiironenHorseshoe}
Piironen, J. and Vehtari, A. (2017).
\newblock \enquote{{Sparsity information and regularization in the horseshoe
  and other shrinkage priors}.}
\newblock {\em Electronic Journal of Statistics\/}, 11(2): 5018 -- 5051.
\newline\urlprefix\url{https://doi.org/10.1214/17-EJS1337SI}
\endbibitem

\bibitem[{Polson and Scott(2011)}]{PolsonLevy}
Polson, N.~G. and Scott, J.~G. (2011).
\newblock \enquote{Local shrinkage rules, Levy processes, and regularized
  regression.}
\endbibitem

\bibitem[{Polson and Scott(2012)}]{PolsonHalfCauchy}
--- (2012).
\newblock \enquote{{On the Half-Cauchy Prior for a Global Scale Parameter}.}
\newblock {\em Bayesian Analysis\/}, 7(4): 887 -- 902.
\newline\urlprefix\url{https://doi.org/10.1214/12-BA730}
\endbibitem

\bibitem[{Polson et~al.(2013)Polson, Scott, and Windle}]{polson2013bayesian}
Polson, N.~G., Scott, J.~G., and Windle, J. (2013).
\newblock \enquote{Bayesian inference for logistic models using
  P{\'o}lya--Gamma latent variables.}
\newblock {\em Journal of the American statistical Association\/}, 108(504):
  1339--1349.
\endbibitem

\bibitem[{Robert(1991)}]{RobertGIG}
Robert, C. (1991).
\newblock \enquote{Generalized inverse normal distributions.}
\newblock {\em Statistics \& Probability Letters\/}, 11(1): 37--41.
\newline\urlprefix\url{https://www.sciencedirect.com/science/article/pii/016771529190174P}
\endbibitem

\bibitem[{Robert(2007)}]{robert2007bayesian}
--- (2007).
\newblock {\em The Bayesian Choice: From Decision-Theoretic Foundations to
  Computational Implementation\/}.
\newblock Springer Texts in Statistics. Springer New York.
\newline\urlprefix\url{https://books.google.ch/books?id=6oQ4s8Pq9pYC}
\endbibitem

\bibitem[{Robert and Casella(2004)}]{robert_monte_2004}
Robert, C.~P. and Casella, G. (2004).
\newblock {\em Monte {Carlo} {Statistical} {Methods}\/}.
\newblock Springer {Texts} in {Statistics}. New York, NY: Springer.
\newline\urlprefix\url{http://link.springer.com/10.1007/978-1-4757-4145-2}
\endbibitem

\bibitem[{Ročkov{\'a}(2018)}]{Rockov2018BayesianEO}
Ročkov{\'a}, V. (2018).
\newblock \enquote{Bayesian estimation of sparse signals with a continuous
  spike-and-slab prior.}
\newblock {\em Annals of Statistics\/}, 46: 401--437.
\newline\urlprefix\url{https://api.semanticscholar.org/CorpusID:85503428}
\endbibitem

\bibitem[{Song and Liang(2017)}]{Song2017NearlyOB}
Song, Q. and Liang, F. (2017).
\newblock \enquote{Nearly optimal Bayesian shrinkage for high-dimensional
  regression.}
\newblock {\em Science China Mathematics\/}, 66: 409--442.
\newline\urlprefix\url{https://api.semanticscholar.org/CorpusID:88516067}
\endbibitem

\bibitem[{{Stan Development Team}(2022)}]{stan2022}
{Stan Development Team} (2022).
\newblock \enquote{Stan Modeling Language Users Guide and Reference Manual,
  Version 2.34.0.}
\newline\urlprefix\url{http://mc-stan.org/}
\endbibitem

\bibitem[{Stein(1981)}]{stein_estimation_1981}
Stein, C.~M. (1981).
\newblock \enquote{Estimation of the {Mean} of a {Multivariate} {Normal}
  {Distribution}.}
\newblock {\em The Annals of Statistics\/}, 9(6): 1135--1151.
\newblock Publisher: Institute of Mathematical Statistics.
\newline\urlprefix\url{https://projecteuclid.org/journals/annals-of-statistics/volume-9/issue-6/Estimation-of-the-Mean-of-a-Multivariate-Normal-Distribution/10.1214/aos/1176345632.full}
\endbibitem

\bibitem[{Stryjewski(2010)}]{boxplots}
Stryjewski, L. (2010).
\newblock \enquote{40 years of boxplots.}
\newline\urlprefix\url{https://api.semanticscholar.org/CorpusID:36975036}
\endbibitem

\bibitem[{Stuart and Ord(2009)}]{stuart2009kendall}
Stuart, A. and Ord, K. (2009).
\newblock {\em Kendall's Advanced Theory of Statistics: Volume 1: Distribution
  Theory\/}.
\newblock Number vol.~1~;vol.~1994 in Kendall's Advanced Theory of Statistics.
  Wiley.
\newline\urlprefix\url{https://books.google.ch/books?id=tW18thQWJQIC}
\endbibitem

\bibitem[{Szakács et~al.(2004)Szakács, Annereau, Lababidi, Shankavaram,
  Arciello, Bussey, Reinhold, Guo, Kruh, Reimers, Weinstein, and
  Gottesman}]{SZAKACS2004129}
Szakács, G., Annereau, J.-P., Lababidi, S., Shankavaram, U., Arciello, A.,
  Bussey, K.~J., Reinhold, W., Guo, Y., Kruh, G.~D., Reimers, M., Weinstein,
  J.~N., and Gottesman, M.~M. (2004).
\newblock \enquote{Predicting drug sensitivity and resistance: Profiling ABC
  transporter genes in cancer cells.}
\newblock {\em Cancer Cell\/}, 6(2): 129--137.
\newline\urlprefix\url{https://www.sciencedirect.com/science/article/pii/S1535610804002065}
\endbibitem

\bibitem[{Tadesse and Vannucci(2021)}]{bayesian_variable_selection_handbook}
Tadesse, M. and Vannucci, M. (2021).
\newblock {\em Handbook of Bayesian Variable Selection\/}.
\newblock Chapman \& Hall/CRC Handbooks of Modern Statistical Methods. CRC
  Press.
\newline\urlprefix\url{https://books.google.de/books?id=Cn1TEAAAQBAJ}
\endbibitem

\bibitem[{Tibshirani(1996)}]{tibshirani1996lasso}
Tibshirani, R. (1996).
\newblock \enquote{{Regression shrinkage and selection via the lasso}.}
\newblock {\em Journal of the Royal Statistical Society. Series B
  (Methodological)\/}, 267--288.
\endbibitem

\bibitem[{Tosh et~al.(2022)Tosh, Greengard, Goodrich, Gelman, Vehtari, and
  Hsu}]{piranha1}
Tosh, C., Greengard, P., Goodrich, B., Gelman, A., Vehtari, A., and Hsu, D.
  (2022).
\newblock \enquote{The piranha problem: Large effects swimming in a small
  pond.}
\endbibitem

\bibitem[{van~der Pas(2021)}]{vanDP2021theoretical}
van~der Pas, S. (2021).
\newblock \enquote{Theoretical guarantees for the horseshoe and other
  global-local shrinkage priors.}
\newblock In {\em Handbook of Bayesian Variable Selection\/}, 133--160. Chapman
  and Hall/CRC.
\endbibitem

\bibitem[{van~der Pas et~al.(2016)van~der Pas, Salomond, and
  Schmidt-Hieber}]{vanDP2016Conditions}
van~der Pas, S., Salomond, J.-B., and Schmidt-Hieber, J. (2016).
\newblock \enquote{{Conditions for posterior contraction in the sparse normal
  means problem}.}
\newblock {\em Electronic Journal of Statistics\/}, 10(1): 976 -- 1000.
\newline\urlprefix\url{https://doi.org/10.1214/16-EJS1130}
\endbibitem

\bibitem[{van~der Pas et~al.(2017)van~der Pas, Szab{\'o}, and van~der
  Vaart}]{vanDP2017UQHorseshoe}
van~der Pas, S., Szab{\'o}, B., and van~der Vaart, A. (2017).
\newblock \enquote{{Uncertainty Quantification for the Horseshoe (with
  Discussion)}.}
\newblock {\em Bayesian Analysis\/}, 12(4): 1221 -- 1274.
\newline\urlprefix\url{https://doi.org/10.1214/17-BA1065}
\endbibitem

\bibitem[{van~der Pas et~al.(2014)van~der Pas, Kleijn, and van~der
  Vaart}]{vanderpasblackhs}
van~der Pas, S.~L., Kleijn, B. J.~K., and van~der Vaart, A.~W. (2014).
\newblock \enquote{{The horseshoe estimator: Posterior concentration around
  nearly black vectors}.}
\newblock {\em Electronic Journal of Statistics\/}, 8(2): 2585 -- 2618.
\newline\urlprefix\url{https://doi.org/10.1214/14-EJS962}
\endbibitem

\bibitem[{{Van Erp} et~al.(2019){Van Erp}, Oberski, and
  Mulder}]{BayesPenalizedRegSara}
{Van Erp}, S., Oberski, D., and Mulder, J. (2019).
\newblock \enquote{Shrinkage priors for Bayesian penalized regression.}
\newblock {\em Journal of Mathematical Psychology\/}, 89: 31--50.
\endbibitem

\bibitem[{Vehtari et~al.(2023)Vehtari, Gabry, Magnusson, Yao, Bürkner,
  Paananen, and Gelman}]{looR}
Vehtari, A., Gabry, J., Magnusson, M., Yao, Y., Bürkner, P.-C., Paananen, T.,
  and Gelman, A. (2023).
\newblock \enquote{loo: Efficient leave-one-out cross-validation and WAIC for
  Bayesian models.}
\newblock R package version 2.6.0.
\newline\urlprefix\url{https://mc-stan.org/loo/}
\endbibitem

\bibitem[{Vehtari et~al.(2016)Vehtari, Gelman, and Gabry}]{VehtariWAIC}
Vehtari, A., Gelman, A., and Gabry, J. (2016).
\newblock \enquote{Practical Bayesian model evaluation using leave-one-out
  cross-validation and WAIC.}
\newblock {\em Statistics and Computing\/}, 27(5): 1413–1432.
\newline\urlprefix\url{http://dx.doi.org/10.1007/s11222-016-9696-4}
\endbibitem

\bibitem[{Vehtari et~al.(2021)Vehtari, Gelman, Simpson, Carpenter, and
  B{\"u}rkner}]{vehtariRhat}
Vehtari, A., Gelman, A., Simpson, D., Carpenter, B., and B{\"u}rkner, P.-C.
  (2021).
\newblock \enquote{{Rank-Normalization, Folding, and Localization: An Improved
  $\widehat{R}$ for Assessing Convergence of MCMC (with Discussion)}.}
\newblock {\em Bayesian Analysis\/}, 16(2): 667 -- 718.
\newline\urlprefix\url{https://doi.org/10.1214/20-BA1221}
\endbibitem

\bibitem[{Vehtari and Ojanen(2012)}]{Vehtari2012Predictive}
Vehtari, A. and Ojanen, J. (2012).
\newblock \enquote{{A survey of Bayesian predictive methods for model
  assessment, selection and comparison}.}
\newblock {\em Statistics Surveys\/}, 6(none): 142 -- 228.
\newline\urlprefix\url{https://doi.org/10.1214/12-SS102}
\endbibitem

\bibitem[{Vehtari et~al.(2022)Vehtari, Simpson, Gelman, Yao, and
  Gabry}]{vehtariPSIS}
Vehtari, A., Simpson, D., Gelman, A., Yao, Y., and Gabry, J. (2022).
\newblock \enquote{Pareto Smoothed Importance Sampling.}
\endbibitem

\bibitem[{Vershynin(2018)}]{vershynin2018high}
Vershynin, R. (2018).
\newblock {\em High-Dimensional Probability: An Introduction with Applications
  in Data Science\/}.
\newblock Cambridge Series in Statistical and Probabilistic Mathematics.
  Cambridge University Press.
\newline\urlprefix\url{https://books.google.de/books?id=J-VjswEACAAJ}
\endbibitem

\bibitem[{Wang and Polson(2024)}]{wang2024pochhammerpriorssparsecount}
Wang, Y. and Polson, N.~G. (2024).
\newblock \enquote{Pochhammer Priors for Sparse Count Models.}
\newline\urlprefix\url{https://arxiv.org/abs/2402.09583}
\endbibitem

\bibitem[{West(1987)}]{mwestScaleNormals}
West, M. (1987).
\newblock \enquote{{On scale mixtures of normal distributions}.}
\newblock {\em Biometrika\/}, 74(3): 646--648.
\newline\urlprefix\url{https://doi.org/10.1093/biomet/74.3.646}
\endbibitem

\bibitem[{Wong(1998)}]{GenDirWONG1998}
Wong, T.-T. (1998).
\newblock \enquote{Generalized Dirichlet distribution in Bayesian analysis.}
\newblock {\em Applied Mathematics and Computation\/}, 97(2): 165--181.
\newline\urlprefix\url{https://www.sciencedirect.com/science/article/pii/S0096300397101400}
\endbibitem

\bibitem[{Xun et~al.(2017)Xun, Li, Zhao, Gao, and Zhang}]{CTMijcai2017p588}
Xun, G., Li, Y., Zhao, W.~X., Gao, J., and Zhang, A. (2017).
\newblock \enquote{A Correlated Topic Model Using Word Embeddings.}
\newblock In {\em Proceedings of the Twenty-Sixth International Joint
  Conference on Artificial Intelligence, {IJCAI-17}\/}, 4207--4213.
\newline\urlprefix\url{https://doi.org/10.24963/ijcai.2017/588}
\endbibitem

\bibitem[{Yang et~al.(2016)Yang, Wainwright, and Jordan}]{BayesianHDComplexity}
Yang, Y., Wainwright, M.~J., and Jordan, M.~I. (2016).
\newblock \enquote{ON THE COMPUTATIONAL COMPLEXITY OF HIGH-DIMENSIONAL BAYESIAN
  VARIABLE SELECTION.}
\newblock {\em The Annals of Statistics\/}, 44(6): 2497--2532.
\newline\urlprefix\url{http://www.jstor.org/stable/44245760}
\endbibitem

\bibitem[{Zhang and Bondell(2018)}]{ZhangCredibleRegions}
Zhang, Y. and Bondell, H.~D. (2018).
\newblock \enquote{{Variable Selection via Penalized Credible Regions with
  Dirichlet–Laplace Global-Local Shrinkage Priors}.}
\newblock {\em Bayesian Analysis\/}, 13(3): 823 -- 844.
\newline\urlprefix\url{https://doi.org/10.1214/17-BA1076}
\endbibitem

\bibitem[{Zhang et~al.(2020)Zhang, Naughton, Bondell, and Reich}]{r2d2zhang}
Zhang, Y.~D., Naughton, B.~P., Bondell, H.~D., and Reich, B.~J. (2020).
\newblock \enquote{Bayesian Regression Using a Prior on the Model Fit: The
  R2-D2 Shrinkage Prior.}
\newblock {\em Journal of the American Statistical Association\/}, 0(0): 1--13.
\newline\urlprefix\url{https://doi.org/10.1080/01621459.2020.1825449}
\endbibitem

\end{thebibliography}

\appendix
\section{Supplementary Results}

\label{appendix}

\subsection{Shifting between Logistic Normal representations}

We have explored different transformations from the real space into the simplex. Namely the additive log-ratio (alr) and the symmetric logistic transformation. The alr transformation relative to the $K$th component is defined as: \citep{logisticnormal}.
%---------------------------------------
\begin{align}
\label{eq:logistic_alr}   
\eta= \alr(\phi) \coloneqq \left(  \ln \left( \frac{\phi_{1}}{\phi_{K}} \right), ...,  \ln \left( \frac{\phi_{K-1}}{\phi_{K}} \right)  \right) \in \mathbb{R}^{K-1}, 
\end{align}
%---------------------------------------
where $\phi \in \mathcal{S}^{K}$. The inverse alr transformation is defined by the\textit{ standard logistic} (\textit{softmax}) transformation with a fill-up term for the reference component: $\phi_k= \frac{ e^{\eta_k}}{1+\sum_{j=1}^{K-1} e^{\eta_j} }, k=1,..., K-1, ,  \phi_K= 1-\sum_{k=1}^{K-1}\phi_k.$ 

If $\eta = \alr(\phi)$ follows a multivariate normal distribution in $\mathbb{R}^{K-1}$, then $\phi$ is said to follow an additive logistic normal (ALN) distribution, denoted as $\phi \sim \text{ALR}(\mu, \Sigma)$. The parameters $\mu$ and $\Sigma$ depend on the chosen reference component. 

The symmetric logistic transformation does not rely on a reference component. Here, we start with an unconstrained vector $\theta \in \mathbb{R}^{K}$, assumed to follow a multivariate normal distribution $\theta \sim \normal_K(\mu, \Sigma)$, and map it onto the simplex via
%---------------------------------------
\begin{align}
\label{eq:logistic_map_app}
\phi_k = \frac{e^{\theta_k}}{\sum_{j=1}^K e^{\theta_j}} , \ k=1, \ldots, K.
\end{align}
%---------------------------------------
In this case, $\phi$ follows a logistic normal (LN) distribution, written as $\phi \sim \logisticnormal(\mu, \Sigma)$. To ensure identifiability, it is common to impose either the constraint $\sum_{k=1}^K \theta_k = 0$ or fix one component, e.g., $\theta_K = 0$. Other options are available as well.  The latter case corresponds to the ALR transformation, while under the sum-to-zero constraint, we can express $\theta_k$ in terms of $\phi$ as $\theta_k = \ln \left( \frac{\phi_k}{g(\phi)} \right)$, where $g(\phi)$ is the geometric mean of $\phi$.

To translate between the two representations, we consider their associated mean and covariance structures. Let $\Sigma = [\sigma_{ij}]$ denote the covariance matrix of the alr-transformed variables,

\begin{align*}
\sigma_{ij} = \cov  \left(  \log \left( \frac{\phi_i}{ \phi_K} \right), \log \left( \frac{\phi_j}{ \phi_K}  \right)  \right), \quad i, j = 1,\dots, K-1.
\end{align*}

and let $\Gamma= [\gamma_{ij}]$ be the covariance matrix of the centered log-transformed variables,

\begin{align*}
\gamma_{ij} =  \cov  \left(  \log \left( \frac{\phi_i}{ g(\phi)} \right), \log \left( \frac{\phi_j}{ g(\phi)} \right)   \right), \quad i, j = 1,\dots, K.
\end{align*}

The expectation and covariance relationships between $\eta$ and $\theta$ are then given by

\begin{align*}
\mbe(\eta_i) &=  \mbe(\theta_i)- \mbe(\theta_K), \quad i = 1, \dots, K-1, \\
\mbe(\theta_i) &=  \frac{K-1}{ K} \mbe(\eta_i) - \sum_{j \neq i} \mbe( \eta_j ), \quad i = 1, \dots, K-1, \\
\sigma_{ij} &= \gamma_{ij} -\gamma_{iK} - \gamma_{jK} + \gamma_{KK}, \ \ i, j= 1,..., K-1\\
\gamma_{ij} &= \sigma_{ij} -\sigma_{i\cdot} - \sigma_{j\cdot} + \sigma_{\cdot \cdot} \ \ i, j= 1,..., K, 
\end{align*}

where the subscript $\cdot$ denotes summation over all possible values of that index, followed by division by $K$. For instance $\sigma_{i.} = K^{-1} \sum_j \sigma_{ij}$ and $\sigma_{..} = K^{-2} \sum_i \sum_j \sigma_{ij}$. 

The change of representation is particularly useful when employing prior matching, where the user first determines $\mu^*, \Sigma^*$ for $\eta$ and then derives the corresponding mean and covariance $\Gamma$ for $\theta$.

\subsection{Taylor approximation for the mean of a ratio}

The following result appears in \cite{stuart2009kendall}. We show it, given it is central to the discussion of hyperparameter specification. 

Let $X $ and $Y$ have support in $(0, \infty)$. For $f(x,y)$, the bivariate first order Taylor expansion around $(a, b)$ is given by 

\begin{align*}
f(x,y) = f(a,b)+ f_x'(a,b)(x-a) + f_y'(a,b)(y-a) + Q 
\end{align*} 

where $f_x, f_y$ denote the partial derivatives and $Q$ is a remainder of order smaller than the other terms. Assume that $\mu_x = \mbe(X),\mu_y = \mbe(Y) $ exist and select the point of expansion as $(\mu_x,\mu_y)$. An approximation to $\mbe(f(X,Y))$ is given by

\begin{align*}
    \mbe\left(f(X,Y)\right) &\approx \mbe\left[ f \left(\mu_x, \mu_y\right) + f'_x \left(\mu_x,\mu_y\right) (X-\mu_x )+
    f'_y \left(\mu_x,\mu_y\right) (Y-\mu_y )\right] \\
    &= \mbe\left[ f \left(\mu_x, \mu_y\right) \right] + f'_x \left(\mu_x,\mu_y\right) \mbe \left[(X-\mu_x ) \right]+
    f'_y \left(\mu_x,\mu_y\right) \mbe \left[ (Y-\mu_y )\right] \\
    &= f(\mu_x, \mu_y).
\end{align*}

If $f(x,y) ) = \frac{x}{y}$, then using $(\mu_x, \mu_y)$ as expansion point, results in $\mbe \left( \frac{X}{Y} \right) = \frac{\mu_x}{\mu_y}$. 

 We can improve the first order approximation. The second order Taylor approximation around $(a,b)$ is given by 

\begin{align*}
f(x,y) &= f(a,b)+ f_x'(a,b)(x-a) + f_y'(a,b)(y-a) + \\& \quad 1/2 \left[  f_{xx}''(x, y) (x- a)^2 +2f_{xy}''(x, y) (x- a)(y-b)+ f_{yy}''(x, y) (y- b)^2 \right] + Q, 
\end{align*} 

where $f_{xx}, f_{yy}$ are the second order partial derivatives and $Q$ is of small order. Choosing $(a,b) = (\mu_x, \mu_y)$ yields

\begin{align*}
\mbe\left(f(X,Y) \right)  & \approx f(\mu_x,\mu_y)+ 1/2 \left[  f_{xx}''(\mu_x, \mu_y) \var(X) +2f_{xy}''(\mu_x, \mu_y) \text{Cov}(X, Y)+ f_{yy}''(\mu_x,\mu_y) \var(Y)   \right]. 
\end{align*} 

If $f(x,y) ) = \frac{x}{y}$, then using $(\mu_x, \mu_y)$ as expansion point, provides us with $\mbe \left( \frac{X}{Y} \right) \approx \frac{\mathbb{E}[X]}{\mathbb{E}[Y]} - \frac{\text{Cov}(X, Y)}{(\mathbb{E}[Y])^2} + \frac{\text{Var}(Y) \cdot \mathbb{E}[X]}{(\mathbb{E}[Y])^3} $. The last expression is not always available in close form and we must resort to Monte Carlo methods.

\subsection{Gibbs with slice sampling}

A Gibbs sampler for the case in which a Dirichlet decomposition is involved has been proposed by \cite{aguilar_intuitive_2023}. We proceed to show a Gibbs sampler with an imbedded slice sampler for the case in which a Logistic Normal decomposition is used \citep{neal_slice_2003, robert_monte_2004}. For ease of notation and to be able to develop a blocked Gibbs sampling we will write the GDR2 model in matrix form. This will drastically improve performance over moving one regression term at a time. 
	
Denote by $b=(b_1,...,b_K)'$ and let $y=(y_1,...,y_N)'$ denote a vector containing the $N$ observations $y_i, i=1,...,N$, $X$ denote the design matrix of dimension $N \times p$.  Let $\Sigma_b=\sigma^2 \Gamma_b$ denote the covariance matrix of the vector of the coefficients $b$ where $\Gamma_b=\text{diag}\left\lbrace \phi_1  \tau^2 ,..., \phi_K  \tau^2    \right\rbrace$. Let $I_N$ represent the identity matrix of order $N$.  With this notation we can write the GDR2 model in matrix form as: 
    \begin{equation}
    \label{eq:gdr2}
    \begin{aligned}
		y & \mid \mu, \sigma^2 \sim \normal(\mu, \sigma^2 I_N), \\  
		b& \mid\sigma, \tau^2, \phi \sim \normal \left(0, \Sigma_b \right), \\ 
		\phi & \sim \text{LN}(\mu, \Sigma) \\
	    \tau^2 \mid \xi &\sim \gammadist (a_1, \xi), \ \ 
		\xi \sim \gammadist(a_2, 1), \ \  \sigma \sim p(\sigma). 
	\end{aligned}
    \end{equation}

We have used the fact that $  \tau^2 \mid \xi \sim \gammadist (a_1, \xi),  \xi \sim \gammadist(a_2, 1)$ is equivalent to $\tau^2 \sim \betaprime (a_1, a_2)$ \citep{r2d2zhang}.  In the following we denote by $Z \sim \gigdist \left(  \chi, \rho, \nu  \right) $, the Generalized Inverse Gaussian distribution  \citep{RobertGIG} with parameters $\chi>0, \rho>0$ and $\nu \in \mathbb{R}$  if 
	\[ p(z) \propto  z^{\nu-1} \exp\left\lbrace   -\left(  \rho z+ \chi/z  \right)/2    \right\rbrace. \] 

We denote by $\invgammadist(c,d)$ the Inverse Gamma distribution with shape $c$ and scale $d$ and consider that a priori  $\sigma^2 \sim \invgammadist(c,d)$. Consider representation \eqref{eq:gdr2} of the GDR2 prior, then the Gibbs sampling procedure is the following: 
	
	\begin{enumerate}
		\item Set initial values for $b, \sigma, \phi, \tau^2, \xi$. 
		\item Sample $b \mid \phi, \tau, \sigma, y  \sim \normal \left(  \bar{b}  , S_{b} \right)$, where 
		
		\begin{align*}
		    \bar{b}= \left(X'X+ \Gamma_b^{-1} \right)^{-1} X'y , \ \
		    S_b= \sigma^2 \left(X'X+ \Gamma_b^{-1} \right)^{-1}. \\
 		\end{align*}
		
		\item Sample $\sigma^2 \mid  b, \phi, \tau^2  \sim \invgammadist \left(  \dot{c}, \dot{d}   \right)$, where 
		
		\begin{align*}
		  \dot{c}=c+\frac{1}{2}\left( N+K \right) , \ \ 
		  \dot{d}=  d+   \frac{1}{2} \left(  ||  y-Xb||_2^2+  b' \Gamma_{b}^{-1} b \right).
		\end{align*}
		
		\item Sample  $\tau^2 \mid b, \sigma, \xi \sim \gigdist  \left( \chi, \rho, \nu \right)$, where
		
		\begin{align*}
		\chi=  \frac{1}{\sigma^2}   b'   T_b^{-1}b , \ \ 
		\rho= 2\xi , \ \ \nu= a_1- K/2,  \\
		\end{align*}
        
        where $T_b$ is such that $\Gamma_b= T_b \tau^2 $. Sampling from the Generalized Inverse Gaussian is not straightforward, but the reader can refer to \cite{RobertGIG,GenInvGaussian} for efficient methods.\\
	
		\item Sample $\xi \mid \tau^2   \sim \gammadist \left( a_1+a_2 , 1+\tau^2 \right) $.\\
		\item Sample $\phi \mid b, u, \sigma, \xi $. \\ 
		
		Instead of sampling from $\phi$, we consider sampling from the vector of log ratios $\eta$. The conditional distribution of $\eta$ is 

        \begin{align}
            p(\eta \mid b, \tau^2, \sigma) \propto \prod_{i=1}^K \exp \left \lbrace     -\frac{b_i^2}{2\tau^2 \sigma^2} (1+ \sum_{j \neq i} \exp(\eta_j))\right \rbrace \exp \left \lbrace     -1/2 (\eta- \mu)' \Sigma^{-1}(\eta- \mu)\right \rbrace.
        \end{align}

        Follow the next steps to use slice sampling \citep{neal_slice_2003, robert_monte_2004}   

        \begin{enumerate}
            \item The log-posterior is:
           \[
           \log p(\eta | b, \tau^2, \sigma) = -\frac{1}{2} (\eta - \mu)' \Sigma^{-1} (\eta - \mu) - \sum_{i=1}^K \frac{b_i^2}{2\tau^2 \sigma^2} \left(1 + \sum_{j \neq i} \exp(\eta_j)\right).
           \]
        
        \item  Use slice sampling for each dimension $\eta_i$. For each dimension of $\eta = (\eta_1, \eta_2, \dots, \eta_K)$ do the following: 
   
        \begin{enumerate}
       \item Introduce an auxiliary variable $u$. Sample $u$ uniformly from $[0, p(\eta \mid b, \tau^2, \sigma)$. This defines the slice: 
       \[
       u \leq \log p(\eta | b, \tau^2, \sigma).
       \]

       \item Create an initial interval $[L, R]$ around the current value of $\eta_i$:
       \[
       L = \eta_i - w, \quad R = \eta_i + w,
       \]
       where $w$ is a tuning parameter.

       \item Propose a new value for $\eta_i$ uniformly from $[L, R]$ and accept it only if $\log p(\eta) \geq u$.
    
       \item  If the proposed value is rejected, shrink the interval $[L, R]$ and repeat until a valid sample is found.

       \item Replace the current $\eta_i$ with the accepted sample and proceed to the next dimension.

      \item  Repeat the process for all dimensions to complete the slice sampling step.
   \end{enumerate}

\end{enumerate}

\item Repeat until convergence 		

It is also possible to create a multivariate slice sampler. See \cite{neal_slice_2003} for details.

\end{enumerate}

\subsection{Stan code}

The Stan codes for the GDR2 prior, with its several decompositions, and the other priors used in the experiments are available at \myosfresults, where it will be maintained and updated as needed.

\section{Further Simulation results}

\subsection{GDR2 models}
 
In the following we show additional simulation results for the simulated and fixed coefficient setup. We show Case (A) of the Simulated coefficient setup, in which each coefficient $b_k$ is generated independently from a Normal distribution with mean 0 and variance $\sigma_b^2= 9$. We also show the case in which the Fixed coefficients are set to a value of 7. For brevity, we show the cases in which $a_\pi = 0.5$. and $(\mu_{R^2}, \varphi_{R^2})= (0.5, 1)$. The data for the other results can be found in \myosfresults.

\subsubsection{Results: Simulated coefficients setup }

Figures \ref{app:fig:delta_lpd_test_sparse_norm_no_cor} and \ref{app:fig:delta_rmse_postbpp_allcases_sparse_norm_no_cor} clearly demonstrate that, in the simpler scenario where no correlation is assumed among the components of $b$, LN decompositions outperform the Dirichlet decomposition. These results are consistent with, and even more promising than, those presented in the main text. However, the correlated case discussed in the main text is more representative of real-world applications and provides a more rigorous test of the priors’ performance.

%----------------------------
\begin{figure}[h]
    \centering
    \includegraphics[width=1\linewidth]{figs/comparison_R2_models/gen_coef_sparse_norm/_delta_lpd_test-gen_coef_sparse_norm.pdf}
    \vspace{-0.5cm}  
    \caption{\textbf{Simulated coefficients setup.} $\Delta \elpd$  evaluated on test datasets of size $N=500$ for the simulated coefficients setup.} 
    \label{app:fig:delta_lpd_test_sparse_norm_no_cor}
\end{figure}
%----------------------------
%----------------------------
\begin{figure}[h]
    \centering
    \includegraphics[width=1\linewidth]%{figs/sims_joint_api_0.5/delta_rmse_postbpp_allcases_id_1}
{figs/comparison_R2_models/gen_coef_sparse_norm/delta_rmse_postbpp_allcases_gen_coef_sparse_norm.pdf}
    \vspace{-0.5cm}  
    \caption{\textbf{Simulated coefficients setup.} Violin plots with embedded box plots for $\Delta \rmse$ under different simulation conditions. $\Delta \rmse$ has been partitioned with respect to truly zero and nonzero coefficients. }
    \label{app:fig:delta_rmse_postbpp_allcases_sparse_norm_no_cor}
\end{figure}
%----------------------------

\subsubsection{Results: Fixed coefficients setup}

The results presented in Figures \ref{app:fig:delta_lpd_test_fixed7} and \ref{app:fig:delta_rmse_postbpp_allcases_fixed7} align with the findings in the main text. In terms of predictive performance, the LNS consistently outperforms all other models. Regarding RMSE, the LN decompositions generally surpass the Dirichlet decompositions, except in the low-signal, high-correlation scenario for the nonzero coefficients. Notably, we observe improved performance when using the LNS instead of the LNF, supporting our earlier conjecture that an optimized choice of hyperparameters could further enhance performance. This also shows that a diagonal structure for $\Sigma$ should serve as a default. 

%----------------------------
\begin{figure}[H]
    \centering
       \includegraphics[width=1\linewidth]{figs/comparison_R2_models/gen_coef_fixed7/_delta_lpd_test-gen_coef_fixed7.pdf}
    \vspace{-0.5cm}  
    \caption{\textbf{Fixed coefficients setup.} $\Delta \elpd$ evaluated on the test datasets for the fixed coefficients when the signals have a value of 7.}
    \label{app:fig:delta_lpd_test_fixed7}
\end{figure}
%---------------------------

%----------------------------
\begin{figure}[H]
    \centering
    \includegraphics[width=1\linewidth]%{figs/sims_joint_api_0.5/delta_rmse_postbpp_allcases_id_1}
{figs/comparison_R2_models/gen_coef_fixed7/delta_rmse_postbpp_allcases_gen_coef_fixed7.pdf}
    \vspace{-0.5cm}  
    \caption{\textbf{Fixed coefficients setup.} Violin plots with embedded box plots for $\Delta \rmse$ under simulation conditions. $\Delta \rmse$ has been partitioned with respect to truly zero and nonzero coefficients. }
    \label{app:fig:delta_rmse_postbpp_allcases_fixed7}
\end{figure}
%----------------------------

\subsubsection{Results: Fixed coefficients setup. Informed scenario.}

In the fixed coefficient setup,  we consider an extra condition, which we have added for the parameter $\alpha$ as follows: $\alpha_k$ takes the value of 10 when the true coefficient $b_k$ is truly nonzero, and 0.5 when $b_k$ is truly equal to zero. This reflects a scenario termed \textit{informative} $\alpha$, where strong user knowledge is incorporated to emphasize important signals. Our primary goals in examining this case are twofold: 1) Assessing the extent to which the Dirichlet and logistic normal decompositions benefit from user-provided information. 2) Investigating whether valuable information is effectively transferred from one prior to another through KL matching. While Dirichlet distributions have been employed in hierarchical priors previously \citep{aguilar_intuitive_2023}, our search did not yield a study where information is directly injected into hyperparameters as demonstrated in this case. Other researchers have highlighted the significance of exploring hyperparameter values that imbue the prior with specific properties \citep{DirichletLaplace, r2d2zhang, aguilar_intuitive_2023}. It is important to note that this informative scenario described here is uncommon in practice, particularly in high-dimensional settings where such detailed prior knowledge is usually unavailable.

\begin{figure}[H]
    \centering
    \includegraphics[width=1\linewidth]{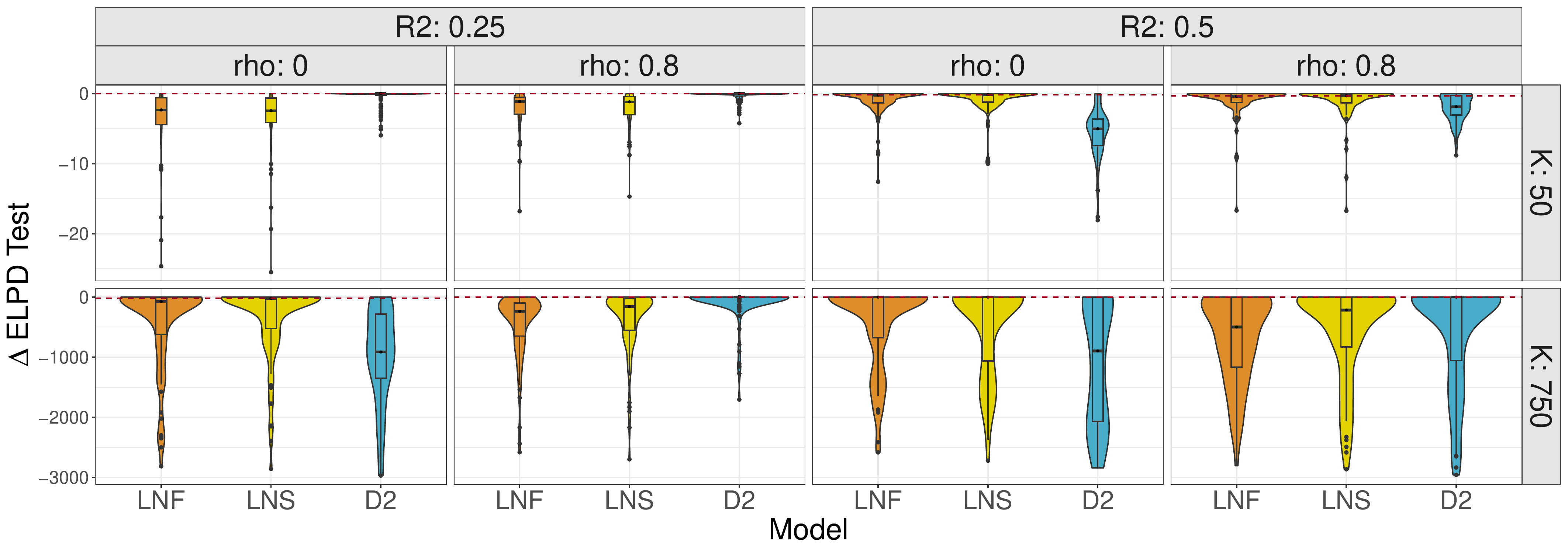}
    \vspace{-0.5cm}  
    \caption{\textbf{Fixed coefficients setup. Informed scenario.} Violin plots with embedded box plots for $\Delta \elpd $ for different simulation conditions.}
    \label{app:fig:delta_lpd_test_id_3}
\end{figure}

\begin{figure}[H]
    \centering
    \includegraphics[width=1\linewidth]{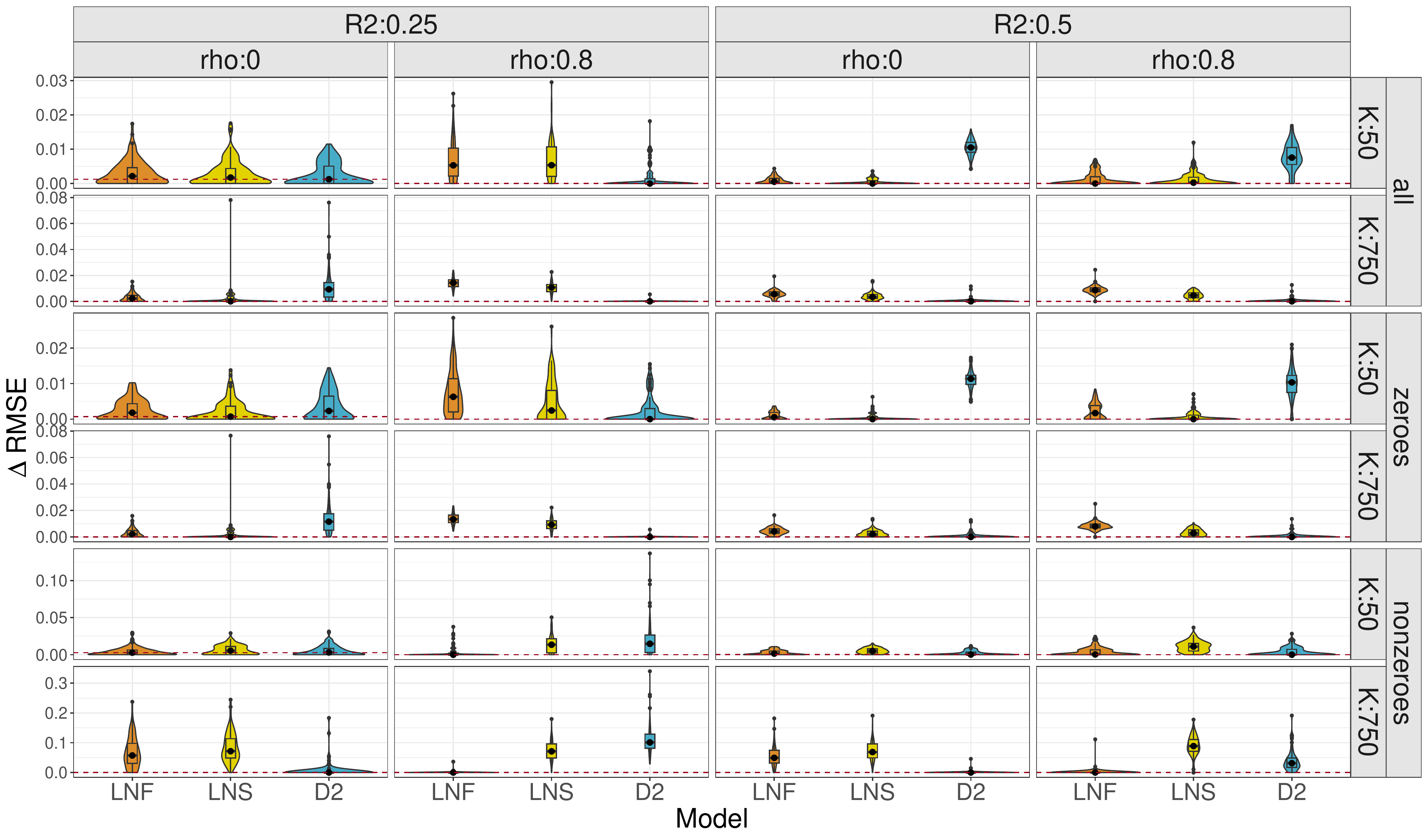}
    \vspace{-0.5cm}  
    \caption{\textbf{Fixed coefficients setup. Informed scenario.} Violin plots with embedded box plots for $\Delta \rmse$ under different simulation conditions. The horizontal line shows the location of the minimum median per condition. We also provide a partition with respect to truly zero and nonzero coefficients. }
    \label{app:fig:delta_rmse_postbpp_allcases_id_3}
\end{figure}

Figure \ref{app:fig:delta_lpd_test_id_3} illustrates how the different models improve their predictive performance when properly guided by the true signals. Similarly, Figure \ref{app:fig:delta_rmse_postbpp_allcases_id_3} displays the distribution of \(\Delta \rmse\) across various simulation conditions for the informed scenario. A closer examination reveals that both LN decompositions consistently outperform the Dirichlet decompositions, with the LNS emerging as the top-performing model.

In general, the three considered priors are becoming more similar, with no clear advantage for either prior. Keep in mind that, in high-dimensional real-world scenarios, users will almost never have this amount of prior knowledge available, rendering the informed prior scenario unrealistic in practice. 

%----------------------------
%----------------------------

%----------------------------
%----------------------------

\subsection{Comparison with other priors}

The shrinkage priors analyzed in this study include the Beta Prime (BP), Dirichlet-Laplace (DL), Horseshoe (HS), and the proposed GDR2 model, which uses Dirichlet and Logistic Normal (LN) decompositions \citep{BaiHypothesisNB, DirichletLaplace, Horseshoe, r2d2zhang}. The Beta Prime, Dirichlet Laplace, and Horseshoe priors were implemented using default hyperparameters, which ensure concentrated mass near the origin and heavy tails. For the Simulated coefficients setup, we present results for a scenario where the coefficients follow an AR(1) correlation structure, defined by $\rho_b = 0.8$, marginal variances $\sigma_b^2 = 9$, and an average sparsity of $\nu = 0.75$. In the Fixed coefficients setup, the coefficients are assigned a constant value of 3. For brevity, we show the cases in which $a_\pi = 0.5$. and $(\mu_{R^2}, \varphi_{R^2})= (0.5, 1)$. Additional results and analyses are available in \myosfresults.

\subsubsection{Results: Simulated coefficients setup}

Figure \ref{app:fig:delta_lpd_test_sparse_norm_cor_all} shows the distribution of $\Delta \elpd$ across simulations under various conditions, visualized as violin plots with embedded boxplots. The results demonstrate that Logistic Normal (LN) decompositions consistently outperform their Dirichlet counterparts. Notably, the LN priors exhibit the best average performance across all considered scenarios. The GDR2 priors perform on par with other well-established shrinkage priors, outperforming them in several cases. In terms of predictive performance, the Beta Prime (BP) and Dirichlet Laplace (DL) priors perform poorly in high-dimensional settings, with values concentrating below $-1e5$ while the classical Horseshoe (HS) often achieves the best performance. Adjusting hyperparameters for $R^2$-based priors might bring them closer to the Horseshoe’s performance; however, our results show that with sensible default settings, they perform competitively.

%----------------------------
\begin{figure}[H]
    \centering
    \includegraphics[width=1\linewidth]{figs/comparison_all_models/gen_coef_sparse_norm_cor/_delta_lpd_test-gen_coef_sparse_norm_cor.pdf}
    \vspace{-0.5cm}  
    \caption{\textbf{Simulated coefficients setup:} $\Delta \elpd$  evaluated on test datasets of size $N=500$ for the simulated coefficients setup.} 
    \label{app:fig:delta_lpd_test_sparse_norm_cor_all}
\end{figure}
%----------------------------

%----------------------------
\begin{figure}[H]
    \centering
    \includegraphics[width=1\linewidth]%{figs/sims_joint_api_0.5/delta_rmse_postbpp_allcases_id_1}
{figs/comparison_all_models/gen_coef_sparse_norm_cor/delta_rmse_postbpp_allcases_gen_coef_sparse_norm_cor.pdf}
    \vspace{-0.5cm}  
    \caption{\textbf{Simulated coefficients setup:} Violin plots with embedded box plots for $\Delta \rmse$ under different simulation conditions. $\Delta \rmse$ has been partitioned with respect to truly zero and nonzero coefficients. }
    \label{app:fig:delta_rmse_postbpp_allcases_sparse_norm_cor}
\end{figure}
%----------------------------

Figure \ref{app:fig:delta_rmse_postbpp_allcases_sparse_norm_cor} depicts the distribution of $\Delta \rmse$ across various simulation conditions. Similar to predictive performance, LN decompositions consistently outperform Dirichlet decompositions in terms of parameter recovery. The LN priors demonstrate the most robust parameter recovery among the competing $R^2$-based models, regardless of the values of $\rho$, $R^2$, or $K$. The overall $\rmse$ results reflect a balance between the model's posterior error for truly zero and nonzero coefficients. To explore this further, Figure \ref{app:fig:delta_rmse_postbpp_allcases_sparse_norm_cor} separately illustrates $\Delta \rmse$ for truly zero and nonzero coefficients.

These results suggest that LN decompositions achieve more effective shrinkage and  compared to Dirichlet counterparts, reducing false positives and improving the detection of truly zero coefficients. The trends for nonzero coefficients are similar. Of particular note is the case with $K = 750$, a challenging scenario where shrinkage priors are expected to overshrink. In this high-dimensional setting, $R^2$-based priors—especially LN decompositions—show improved behavior when moving from $\rho = 0$ to $\rho = 0.80$, performing on par even with the HS. This improvement is also present when transitioning from a low signal ($R^2 = 0.25$) to a moderate signal ($R^2 = 0.60$). The LN decomposition priors perform on par and improving over the performance of well established shrinkage priors.

We present coverage properties using $95\%$ marginal credibility intervals in Table \ref{tab:coverage_sparse_norm_cor} for the case where $\rho = 0$ and $R^2 = 0.6$. Other results can be found in our repository \myosfresults. The table reports the proportion of coverage, specificity, sensitivity, and the average width of the intervals. While frequentist properties are not guaranteed in a Bayesian framework, all models achieve the nominal $95\%$ coverage when $(K, N) = (50, 100)$ and $K = 150$, the latter case being impressive since we are in the high-dimensional setting. However, as $K$ increases to 750, coverage declines due to the increased complexity of the problem.

Specificity and sensitivity remain high for $K = 50, 150$, but for $K = 750$, maintaining adequate sensitivity requires wider intervals, indicating greater uncertainty. Notably, the HS and LN models produce the shortest intervals in this setting. As expected, sensitivity declines sharply due to the increased interval width. Therefore, we recommend supplementing model fit with a second-stage procedure, such as Projection Prediction \cite{PiironenProjInf}, for variable selection rather than relying solely on credibility intervals.

We present ROC curves in Figure \ref{app:fig:roc_coef_sparse_norm_cor}, where the diagonal represents the identity line. The LN models perform comparably to the Horseshoe. In the complex setting with $K = 750$ and $\rho = 0$, the Horseshoe generally outperforms other models. However, when correlation is introduced in $X$, the LNS decomposition surpasses all alternatives.

\begin{table}[!h]
\centering
\caption{\textbf{Simulated coefficients setup:} Coverage properties of $95\%$ marginal posterior credible intervals for the scenario $\rho = 0$, $R^2 = 0.6$}
\label{tab:coverage_sparse_norm_cor}
\resizebox{\ifdim\width>\linewidth\linewidth\else\width\fi}{!}{
\begin{tabular}[t]{cccccc}
\toprule
$K$ & Model ID & Coverage & Specificity & Sensitivity (Power) & Avg. CI Width\\
\midrule
\multirow{6}{*}{50}  & BP  & 0.946 & 0.975 & 0.926 & 0.396\\
                     & DL  & 0.963 & 0.971 & 0.936 & 0.410\\
                     & HS  & 0.974 & 0.985 & 0.931 & 0.382\\
                     & D2  & 0.953 & 0.981 & 0.921 & 0.426\\
                     & LNF & 0.958 & 0.988 & 0.923 & 0.406\\
                     & LNS & 0.963 & 0.994 & 0.919 & 0.389\\
\midrule
\multirow{6}{*}{150} & BP  & 0.978 & 0.997 & 0.707 & 1.500\\
                     & DL  & 0.993 & 1.000 & 0.592 & 2.212\\
                     & HS  & 0.971 & 0.987 & 0.758 & 1.189\\
                     & D2  & 0.985 & 1.000 & 0.584 & 2.192\\
                     & LNF & 0.982 & 0.999 & 0.698 & 1.536\\
                     & LNS & 0.978 & 0.998 & 0.748 & 1.205\\
\midrule
\multirow{6}{*}{750} & BP  & 0.932 & 1.000 & 0.001 & 5.883\\
                     & DL  & 0.943 & 1.000 & 0.001 & 6.137\\
                     & HS  & 0.815 & 1.000 & 0.001 & 2.153\\
                     & D2  & 0.913 & 1.000 & 0.000 & 4.967\\
                     & LNF & 0.890 & 1.000 & 0.000 & 4.311\\
                     & LNS & 0.857 & 1.000 & 0.000 & 3.304\\
\bottomrule
\end{tabular}}
\end{table}

%----------------------------
\begin{figure}[H]
    \centering
    \includegraphics[width=1\linewidth]
{figs/comparison_all_models/gen_coef_sparse_norm_cor/roc_gen_coef_sparse_norm_cor.pdf}
    \vspace{-0.5cm}  
    \caption{\textbf{Simulated coefficients setup:} ROC curves for different models under different simulation conditions. Points are obtained by varying the credibility level of marginal posterior credibility intervals. The grey dotted line depicts the identity line. }
    \label{app:fig:roc_coef_sparse_norm_cor}
\end{figure}
%----------------------------

Figures \ref{app:fig:mcmc_rhat_allcases_sparse_norm_cor}, \ref{app:fig:mcmc_ESS_allcases_sparse_norm_cor}, and \ref{app:fig:mcmc_time_allcases_sparse_norm_cor} present the Effective Sample Size (ESS), Rhat, and total runtime results across the simulations for the coefficient chains, respectively \citep{vehtariRhat}. The Rhat diagnostic indicates no significant convergence issues for most models, with the exception of the Horseshoe, where Rhat values deviate furthest from 1.0. ESS values are generally within acceptable ranges, suggesting efficient sampling across models, although the Horseshoe exhibits the lowest ESS, particularly as model complexity increases. Regarding runtime, Logistic-Normal (LN) models are the fastest for low-complexity cases; however, their computational time grows in higher-complexity scenarios. Finally, the effect of correlations is evident in the runtime distributions, with a noticeable increase in their medians as correlation strength increases.

%----------------------------
\begin{figure}[H]
    \centering
    \includegraphics[width=1\linewidth]
{figs/comparison_all_models/gen_coef_sparse_norm_cor/mcmc_rhat-gen_coef_sparse_norm_cor}
    \vspace{-0.5cm}  
    \caption{\textbf{Simulated coefficients setup:} Distribution of Rhat across different simulations.}
    \label{app:fig:mcmc_rhat_allcases_sparse_norm_cor}
\end{figure}
%----------------------------

%----------------------------
\begin{figure}[H]
    \centering
    \includegraphics[width=1\linewidth]
{figs/comparison_all_models/gen_coef_sparse_norm_cor/mcmc_ess_basic-gen_coef_sparse_norm_cor}
    \vspace{-0.5cm}  
    \caption{\textbf{Simulated coefficients setup:} Distribution of Effective Sample Size across different simulations.}
    \label{app:fig:mcmc_ESS_allcases_sparse_norm_cor}
\end{figure}
%----------------------------

%----------------------------
\begin{figure}[H]
    \centering
    \includegraphics[width=1\linewidth]
{figs/comparison_all_models/gen_coef_sparse_norm_cor/HMC_time-gen_coef_sparse_norm_cor}
    \vspace{-0.5cm}  
    \caption{\textbf{Simulated coefficients setup:} Distribution of total sampling time for each model.}
    \label{app:fig:mcmc_time_allcases_sparse_norm_cor}
\end{figure}
%----------------------------
\subsubsection{Results: Fixed coefficients setup}

Figure \ref{app:fig:delta_lpd_test_fixed3_all} reveals that the LN decompositions are able to perform at par in out-of-sample predictive performance whenever $X$ is uncorrelated and to outperform them  whenever $X$ is highly correlated. Among the R2-based priors, LNS stands out as the most effective predictive model across scenarios. The most pronounced disparities emerge in scenarios characterized by high correlation and dimensionality ($\rho= 0.8, K=750$), where the performance of the Dirichlet prior is notably subpar. The values of BP are extremeley low, therefore not being considered in the plot.
%----------------------------
\begin{figure}[H]
    \centering
    \includegraphics[width=1\linewidth]{figs/comparison_all_models/gen_coef_fixed3/_delta_lpd_test-gen_coef_fixed3.pdf}
    \vspace{-0.5cm}  
    \caption{\textbf{Fixed coefficients setup:} $\Delta \elpd$ evaluated on test datasets of size $N=500$ for the simulated coefficients setup.} 
    \label{app:fig:delta_lpd_test_fixed3_all}
\end{figure}
%----------------------------
The results for $\Delta \rmse$ for all of the coefficients and the zero coefficients are very similar to the simulated coefficients setup. The LN decompositions perform clearly and uniformly better the Dirichlet for all the coefficients and for the zero coefficient only. This indicates again that LN decompositions are better at both overall $\rmse$ and noise detection. 

For $\Delta \rmse$ of nonzero coefficients, the Horseshoe tends to outperform other models across most scenarios, noenetheless LN decompositions perform on par. Among the $R^2$-based priors, no single model consistently dominates. The Dirichlet decomposition excels in finding nonzero coefficients in the in low-signal, high-correlation settings. LN decompositions achieves better performance in all other scenarios. These results suggest that the effectiveness of LN decompositions could be enhanced even more by incorporating covariance matrices that better reflect the dependency structures in $X$. Additionally, alternative hyperparameter settings for $R^2$-based priors might improve their performance, potentially closing the gap with the Horseshoe.

If we focus on the $\rho= 0.8, K=750$ case in both Figures \ref{app:fig:delta_lpd_test_fixed3_all} and \ref{app:fig:delta_rmse_postbpp_fixed3_all}, we can see that there is a tradeoff between out-of-sample predictive performance and detecting signals. Here, the Dirichlet is not shrinking sufficiently (leading to worse out-of-sample predictions), while the LN might be overshrinking some signals (but ensures better out-of-sample predictions overall).
%----------------------------
\begin{figure}[ht]
    \centering
       \includegraphics[width=1\linewidth]{figs/comparison_all_models/gen_coef_fixed3/delta_rmse_postbpp_allcases_gen_coef_fixed3.pdf}
    \vspace{-0.5cm}  
    \caption{\textbf{Fixed coefficients setup} The true value of the signal is 3. $\Delta \rmse$ has been partitioned with respect to truly zero and nonzero coefficients. }
    \label{app:fig:delta_rmse_postbpp_fixed3_all}
\end{figure}
%---------------------------
We also show coverage properties using $95\%$ marginal credibility intervals in Table \ref{tab:coverage_fixed3} for the case where $\rho = 0$ and $R^2 = 0.6$. While frequentist properties are not guaranteed in a Bayesian framework, all models achieve the nominal $95\%$ coverage independently of the value of $K$. 

Specificity and sensitivity remain high for all values of $K$. However, for $K = 750$, models using the Dirichlet distribution (D2, DL) produce wider posterior credibility intervals, which negatively impacts sensitivity. Nevertheless, we still recommend a two-step procedure if selection is sought after.

We present ROC curves in Figure \ref{app:fig:roc_coef_fixed3}. The diagonal line shows the identity line. The LN models perform comparably to the Horseshoe and BP models, even surpassing the latter in some cases. In contrast, models using the Dirichlet distribution tend to underperform, likely due to their wider credibility intervals.

\begin{table}[!h]
\centering
\caption{\textbf{Fixed coefficients setup:} Coverage properties of $95\%$ marginal posterior credible intervals for the scenario $\rho = 0$, $R^2 = 0.6$}
\label{tab:coverage_fixed3}
\resizebox{\ifdim\width>\linewidth\linewidth\else\width\fi}{!}{
\begin{tabular}[t]{cccccc}
\toprule
$K$ & Model ID & Coverage & Specificity & Sensitivity (Power) & Avg. CI Width\\
\midrule
\multirow{6}{*}{50}  & BP  & 0.956 & 0.982 & 0.986 & 0.369\\
                     & DL  & 0.971 & 0.976 & 0.997 & 0.387\\
                     & HS  & 0.987 & 0.996 & 0.997 & 0.351\\
                     & D2  & 0.960 & 0.983 & 0.986 & 0.398\\
                     & LNF & 0.965 & 0.991 & 0.986 & 0.375\\
                     & LNS & 0.970 & 0.996 & 0.987 & 0.358\\
\midrule
\multirow{6}{*}{150} & BP  & 0.926 & 0.941 & 0.980 & 0.375\\
                     & DL  & 0.999 & 1.000 & 0.995 & 0.669\\
                     & HS  & 0.988 & 0.991 & 0.997 & 0.353\\
                     & D2  & 0.991 & 0.999 & 0.975 & 0.647\\
                     & LNF & 0.988 & 0.997 & 0.985 & 0.404\\
                     & LNS & 0.990 & 0.999 & 0.987 & 0.375\\
\midrule
\multirow{6}{*}{750} & BP  & 0.998 & 1.000 & 0.989 & 0.375\\
                     & DL  & 0.989 & 1.000 & 0.222 & 2.585\\
                     & HS  & 0.971 & 0.972 & 0.998 & 0.341\\
                     & D2  & 0.987 & 1.000 & 0.078 & 2.409\\
                     & LNF & 0.998 & 0.999 & 0.988 & 0.441\\
                     & LNS & 0.998 & 1.000 & 0.989 & 0.388\\
\bottomrule
\end{tabular}}
\end{table}
%----------------------------
\begin{figure}[H]
    \centering
    \includegraphics[width=1\linewidth]
{figs/comparison_all_models/gen_coef_fixed3/roc_gen_coef_fixed3.pdf}
    \vspace{-0.5cm}  
    \caption{\textbf{Fixed coefficients setup:} ROC curves for different models under different simulation conditions. Points are obtained by varying the credibility level of marginal posterior credibility intervals. The grey dotted line depicts the identity line. }
    \label{app:fig:roc_coef_fixed3}
\end{figure}
%----------------------------
We also provide MCMC diagnostics. Figures \ref{app:fig:mcmc_rhat_allcases_coef_fixed3}, \ref{app:fig:mcmc_ESS_allcases_coef_fixed3}, and \ref{app:fig:mcmc_time_allcases_coef_fixed3} illustrate the Effective Sample Size (ESS), Rhat, and total runtime results for the coefficient chains across the simulations \citep{vehtariRhat}. The Rhat diagnostic shows no major convergence issues for most models, with the exception of the Horseshoe and Beta Prime, which display the largest deviations from 1.0. ESS values are generally within acceptable ranges, indicating efficient sampling, although the Beta Prime demonstrates the lowest ESS in high-dimensional settings. In terms of runtime, LN models remain the fastest for low-complexity cases. As dimensionality increases, runtime slows across all models, though the increase in time is manageable and does not render any model impractical.
%----------------------------
\begin{figure}[H]
    \centering
    \includegraphics[width=1\linewidth]
{figs/comparison_all_models/gen_coef_fixed3/mcmc_rhat-gen_coef_fixed3}
    \vspace{-0.5cm}  
    \caption{\textbf{Fixed coefficients setup:} Distribution of Rhat across different simulations.}
    \label{app:fig:mcmc_rhat_allcases_coef_fixed3}
\end{figure}
%----------------------------
%----------------------------
\begin{figure}[H]
    \centering
    \includegraphics[width=1\linewidth]
{figs/comparison_all_models/gen_coef_fixed3/mcmc_ess_basic-gen_coef_fixed3}
    \vspace{-0.5cm}  
    \caption{\textbf{Fixed coefficients setup:} Distribution of Effective Sample Size across different simulations.}
    \label{app:fig:mcmc_ESS_allcases_coef_fixed3}
\end{figure}
%----------------------------

%----------------------------
\begin{figure}[H]
    \centering
    \includegraphics[width=1\linewidth]
{figs/comparison_all_models/gen_coef_fixed3/HMC_time-gen_coef_fixed3}
    \vspace{-0.5cm}  
    \caption{\textbf{Fixed coefficients setup:} Distribution of total sampling time for each model.}
    \label{app:fig:mcmc_time_allcases_coef_fixed3}
\end{figure}
%----------------------------

\end{document}